\documentclass{article}
\usepackage[utf8]{inputenc}
\usepackage{amssymb}
\usepackage{geometry}
\usepackage[usenames,dvipsnames]{color}
\usepackage{bbold}
\usepackage{slashed}
\usepackage[table]{xcolor}
\usepackage{graphicx}
\usepackage{mathtools}
\usepackage{svg}
\svgpath{{./Images/}} 

\usepackage{todonotes}
\usepackage{comment}
\usepackage{subfloat}
\usepackage{subcaption}
\usepackage{tabularx}
\usepackage{float}
\usepackage{wrapfig}
\usepackage{booktabs}
\usepackage{hyperref}
\usepackage{amsthm}
\usepackage{pgfplots}
\usepackage[
backend=biber,
style=alphabetic,
sorting=ynt
]{biblatex}
\pgfplotstableread{
x         y    y-max  y-min
Lighting  0.12 0.31   0.03
Computers 0.06 0.12   0.01 
Total     0.07 0.14   0.02
}{\mytable}
\usepackage{algorithm,algpseudocode}
\algdef{SE}[SUBALG]{Indent}{EndIndent}{}{\algorithmicend\ }%
\algtext*{Indent}
\algtext*{EndIndent}
\renewcommand{\Comment}[2][.5\linewidth]{%
  \leavevmode\hfill\makebox[#1][l]{//~#2}}
  
\newtheorem{theorem}{Theorem}[section]
\newtheorem{remark}{Remark}[section]

\DeclarePairedDelimiter\bra{\langle}{\rvert}
\DeclarePairedDelimiter\ket{\lvert}{\rangle}
\DeclarePairedDelimiterX\braket[2]{\langle}{\rangle}{#1 \delimsize\vert #2}
\newcommand{\footremember}[2]{%
\footnote{#2}
\newcounter{#1}
\setcounter{#1}{\value{footnote}}%
}
\newcommand{\footrecall}[1]{%
\footnotemark[\value{#1}]%
}

\graphicspath{{./Images/}}
\title{Real Quantum Amplitude Estimation}
\author{Alberto Manzano$^{a}$\footremember{email}{\textit{Corresponding authors:} \href{mailto:mussodaniele@uniovi.es}{mussodaniele@uniovi.es}  \textit{and} \href{mailto:alberto.manzano.herrero@udc.es}{alberto.manzano.herrero@udc.es}}\footremember{contributions}{\textit{These authors contribute equally to this work.}}, 
Daniele Musso$^{b,c}$\footrecall{email} \footrecall{contributions},
Álvaro Leitao$^{a}$.
}

\date{}

\begin{document}

\maketitle
\begin{center}\it{
$^{a}$Department of Mathematics and CITIC,\\ Universidade da Coruña, A Coruña, Spain\\
\vspace{15pt}
$^{b}$Department of Physics and Instituto de Ciencias y Tecnolog\'ias Espaciales de Asturias (ICTEA),\\
Universidad de Oviedo, Oviedo, Spain\\
\vspace{15pt}
$^{c}$Centro de Supercomputación de Galicia (CESGA),\\ Santiago de Compostela, Spain
}
\end{center}
\vspace{25pt}
\begin{abstract}
We introduce the Real Quantum Amplitude Estimation (RQAE) algorithm, an extension of Quantum Amplitude Estimation (QAE) which is sensitive to the sign of the amplitude. 
RQAE is an iterative algorithm which offers explicit control over the amplification policy through an adjustable parameter.
We provide a rigorous analysis of the RQAE performance and prove that it achieves a quadratic speedup, modulo logarithmic corrections, with respect to unamplified sampling. Besides, we corroborate the theoretical analysis with a set of numerical experiments.
\end{abstract}
\newpage

\tableofcontents

\section{Introduction, motivation and main results}

Quantum Amplitude Estimation (QAE) is an algorithm which retrieves information stored in the amplitude of a quantum state. It is argued to have a quadratic speedup over simple repeated sampling of the quantum state. For this reason, QAE is a central subroutine in quantum computation for various applications, \emph{e.g.}, in chemistry \cite{knill2007optimal,kassal2008polynomial}, finance \cite{rebentrost2018quantum,woerner2019quantum,REV}, and machine learning \cite{wiebe2014quantum,wiebe2016quantum}. The original QAE algorithm \cite{Brassard_2002} is built composing Quantum Phase Estimation (QPE) \cite{nielsen2001quantum} and Grover’s algorithms \cite{grover1996fast}. Standard QPE relies on a Quantum Fourier Transform (QFT) which is very demanding in terms of computational resources, especially if considered for the Noise Intermediate-Scale Quantum era (NISQ). 

Several approaches have been proposed to reduce the resources needed by QAE, both in terms of qubits and circuit depth, while approximately preserving the same speedup. These approaches can be broadly categorized in three families.

\begin{enumerate}
\item
The first family consists in techniques which take advantage of classical post-processing. As an example, in \cite{Suzuki_2020} the authors show how to replace QPE by a set of Grover iterations combined with a Maximum Likelihood Estimation (MLE) post-processing algorithm. To correctly asses the overall performance of such techniques, one needs to include the overhead due to the classical post-processing in the total cost of the algorithm, therefore diminishing the potential speedup. Moreover, at the time of writing, no rigorous proof of the correctness of the proposed algorithms has been given yet.
\item
The second family includes strategies which still rely on phase estimation, but eliminate the need of a QFT. The main idea is to replace the QFT with  Hadamard tests \cite{wie2019simpler}. This variation of QPE was first suggested by Kitaev \cite{kitaev1995quantum} and is called “iterative phase estimation”. In papers following this approach such as \cite{wie2019simpler} it is not even clear how to control the accuracy of the algorithm other than possibly increasing the number of measurements. Besides, they do not give rigorous proof of the correctness of the algorithm.
\item
The methods belonging to the third and last family are based entirely on Grover iterations and they do not require any post-processing. The main difference among algorithms of this family is in the amplification policy. Representative examples of this approach are the Iterative Quantum Amplitude Estimation (IQAE) and the Quantum Amplitude Estimation Simplified (QAES) algorithms \cite{Grinko_2021,Aaronson_2020}. Both provide rigorous proofs of the correctness of the techniques. Although the strategy described in \cite{Aaronson_2020} achieves the desired asymptotic complexity exactly (\emph{i.e.} without logarithmic factors), the constants involved are very large, and likely to render the algorithm impractical. In \cite{Grinko_2021} they do not exactly match the desired asymptotic complexity, yet the constants involved are much lower.
\end{enumerate}

RQAE can be thought of as a generalization of the Quantum Coin algorithm \cite{Abrams1999FastQA,shimada2020quantum} and it is based on an iterative strategy, like \cite{Grinko_2021,Aaronson_2020}. In particular, RQAE utilizes a set of auxiliary amplitudes which allow to shift in a controlled fashion the amplitude to be retrieved. Such shift can be easily and efficiently implemented following the methods presented in \cite{QNP}. Relying on this, we propose a specific strategy to iteratively choose the amplification factor $k$ (\emph{i.e.} the Grover exponent) and the shift $b$ at each iteration, progressively improving the estimation of the quantum amplitude to be retrieved \emph{i.e.} the target amplitude.

We prove for RQAE a set of tight bounds. Moreover, the bounds for RQAE depend on a free parameter $q$ which directly controls the amplification policy. More specifically, the parameter $q$ is a minimum bound for the ratio between the amplification on consecutive steps:
\begin{equation}\label{eq:qqq}
  q\leq q_i 
  \equiv \dfrac{K_{i+1}}{K_i}
  \equiv \dfrac{2k_{i+1}+1}{2k_{i}+1}\ . 
 \end{equation}
 The parameter $q$ affects both the depth of the circuit and the performance (in terms of calls to the oracle) of the algorithm, thus offering a handle to discuss the trade-off between the two.

The other feature that makes RQAE different from alternative amplitude estimation algorithms is the possibility of recovering the sign of the amplitude to be retrieved, hence the name Real Quantum Amplitude Estimation (RQAE). Concretely, RQAE is a \emph{parametric} algorithm that depends on a real input amplitude $b_1$, 
which provides a reference, through which we can unambiguously assign a phase to every other amplitude in the quantum register. Then, when referring to the sign of an amplitude, we mean the relative phase between such amplitude and $b_1$. As the notation already suggests, $b_1$ is the shift mentioned above for the first iteration. The new sensitivity to the relative sign of an amplitude allows one to tackle a wider variety of problems, precluded to standard algorithms.

The remainder of this paper is organized as follows. Section \ref{sec:algorithm} introduces the intuition behind the construction of the algorithm. In Section \ref{sec:configuration_properties} we state some theoretical results on the performance of the algorithm for a specific set of parameters (for the rigorous proof see Appendix \ref{sec:proof}). Moreover, we confirm the theoretical properties with a set of simulated experiments. To conclude, we discuss the results and related open questions in Section \ref{sec:conclusion}.

\section{Real Quantum Amplitude Estimation}\label{sec:algorithm}

\begin{figure}[H]
        \centering
    \begin{subfigure}[t]{0.3\textwidth}
        \centering
        \includegraphics[width=\textwidth]{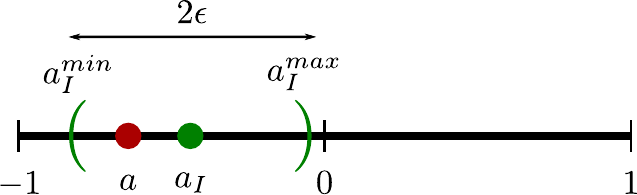}
    \end{subfigure}
     \hfill
    \begin{subfigure}[t]{0.3\textwidth}
        \centering
        \includegraphics[width=\textwidth]{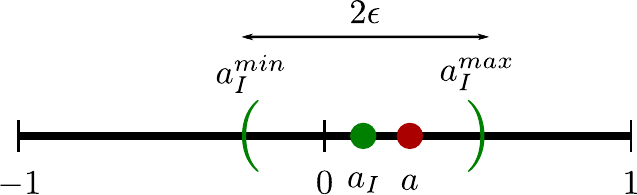}
    \end{subfigure}
     \hfill
    \begin{subfigure}[t]{0.3\textwidth}
        \centering
        \includegraphics[width=\textwidth]{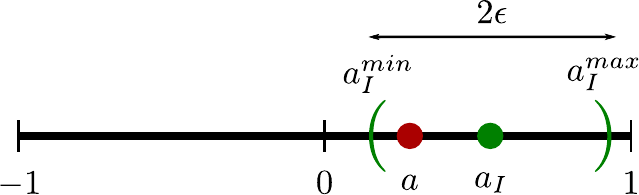}
    \end{subfigure}
    \caption{$a$ is the target amplitude to be estimated, $2\epsilon$ is the width of the estimation interval with bounds $(a_I^{\min},a_I^{\max})$, $a_I$ is the center of the confidence interval. In the image we distinguish three possibilities to emphasize that the algorithm described below is sensitive to the \emph{sign} of $a$.}
    \label{fig:initial_problem}
\end{figure}

Consider a one-parameter family of oracles $\mathcal{A}_{b}$ that, acting on the state $\ket{0}$, yield
\begin{equation}\label{eq:shi_def}
    \mathcal{A}_{b}\ket{0} = \ket{\psi} = \left(a+b\right)\ket{\phi}+c_b\, \ket{\phi^\perp}_b\ ,
\end{equation}
where $a$ is a real number, $b$ is an auxiliary, continuous and real parameter that we call ``shift'', and $\ket{\phi}$ is a specified direction in the Hilbert space ${\cal H}$.
The RQAE algorithm estimates the amplitude $a$ exploiting the possibility of tuning the shift $b$ iteratively. The ket $\ket{\psi}$ belongs to the plane $\Pi_b = \text{span}\{\ket{\phi},\ket{\phi^\perp}_b\} \subset {\cal H}$ for which the kets $\ket{\phi}$ and $\ket{\phi^\perp}_b$ provide an orthonormal basis. Note that all the quantities with a sub-index $b$ depend on the actual value of the shift. In practice, the construction of oracles such as $\mathcal{A}_b$ from a given un-shifted oracle $\mathcal{A}_0$ is generally not difficult. In most cases, a controlled shift of an amplitude can be efficiently implemented via Hadamard gates and some controlled operations. We give an example on how to build such a shifted oracle in Appendix \ref{sec:shift}. In particular, its implementation is straightforward in the framework described in \cite{QNP}.\\ \\ 
Given a precision level $\epsilon$ and a confidence level $1-\gamma$, the goal of the algorithm is to compute an interval $(a_I^{\min},a_I^{\max}) \subset [-1,1]$ of width smaller than $2\epsilon$ which contains the value of $a$ with probability greater or equal to $1-\gamma$ (see Figure \ref{fig:initial_problem}). We take as a representative of the interval its center, $a_I = \frac{a_I^{\min}+a_I^{\max}}{2}$, thus admitting a maximum error of $\epsilon$:
\begin{equation}\label{eq:initial_problem}
        \mathbb{P}\Big[|a-a_I|\geq \epsilon \Big] \leq \gamma\ .
\end{equation}

It is convenient to express the amplitudes in terms of their corresponding angles, that is, we consider the generic mapping $\theta_x = \arcsin(x)$ for any real amplitude $x$. Note that the angle representation is particularly suited to describe Grover amplifications, which indeed admit an interpretation as rotations in the plane $\Pi_b$. As an example, the state $\ket{\psi}$ given in \eqref{eq:shi_def} can be written as 
\begin{equation}
    \ket{\psi} = \sin(\theta_{a+b})\ket{\phi}+\cos(\theta_{a+b})\ket{\phi^\perp}_b\ ,
\end{equation}
where $\theta_{a+b}$ represents a rotation in the plane $\Pi_b$ defined above.
Throughout the paper, we will be changing back and forth from the representation in terms of the actual amplitude or its associated angle whenever needed. To avoid notational clutter, we henceforth drop the sub-index $b$ on the perpendicular ket, leaving its dependence on the shift as understood. Actually, such dependence does not play any role for the algorithm.
\\

In the following subsections we address the details of the procedure describing all the steps contained in each iteration.

\subsection{First iteration: estimating the sign}\label{sign}
This step achieves a first estimation of the bounds of the confidence interval $(a_1^{\min},a_1^{\max})$.  Normally, this estimation would not be sensitive to the sign of the underlying amplitude because, when sampling from a quantum state, we obtain the square of the amplitude. Nevertheless, taking advantage of the shift $b$ we can circumvent this limitation. In order to compute the sign, we will combine two different pieces of information: the result of measuring the two oppositely shifted states $\ket{\psi_1}_{\pm}$ defined as:
\begin{equation}
\begin{aligned}
    \ket{\psi_1}_{+}
    &:=
    \left(a+b_1\right)\ket{\phi} + ... \ ,\\ 
    \ket{\psi_1}_{-}
    &:= \left(a-b_1\right) \ket{\phi} + ...\ ,
\end{aligned}
\end{equation}
for an arbitrary real shift $b_1$. The sign of $b_1$ has to be decided at the start of the algorithm to have a clear reference. In practice, in some setups it is possible to measure at the same time both states taking advantages of Hadamard gates as in the quantum arithmetic techniques discussed in \cite{QNP} (more details are given in Appendix \ref{sec:shift}).
As $a$ and $b_1$ are real numbers, we have the identity: 
\begin{equation}\label{ide}
     a = \dfrac{\left(a+b_1\right)^2-\left(a-b_1\right)^2}{4b_1}\ ,
\end{equation}
and we can build a first empirical estimation $\hat{a}_1$ of $a$ as follows:
\begin{equation}\label{eqn:first_empirical_amplitude_estimation}
    \hat{a}_1 = \dfrac{\hat{p}_{\text{sum}}-\hat{p}_{\text{diff}}}{4b_1}\ ,
\end{equation}
where $\hat{p}_{\text{sum}}$ and $\hat{p}_{\text{diff}}$ are the empirical probabilities of getting $\ket{\phi}$ when measuring $\ket{\psi_1}_{-}$ and $\ket{\psi_1}_{+}$, respectively. Throughout the paper, when we measure, we will use $\hat{p}$ to denote the empirical probability obtained from direct sampling. As an example, if in iteration $i$ we sample the state $N_i$ times, getting $\ket{\phi}$ as an outcome $\hat{N}_i$ times, the estimated probability of $\ket{\phi}$ will be $\hat{p}_i = \dfrac{\hat{N}_i}{N_i}$.\\

\begin{figure}[H]
        \centering
    \begin{subfigure}[c]{0.45\textwidth}
        \centering
        \includegraphics[width=\textwidth]{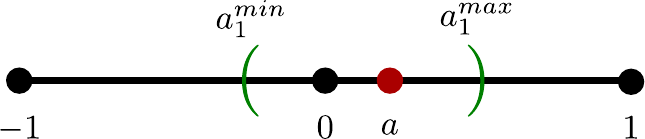}
        \label{fig:amplitude_problem_image}
    \end{subfigure}
        \hfill
    \begin{subfigure}[c]{0.45\textwidth}
        \centering
        \includegraphics[width=\textwidth]{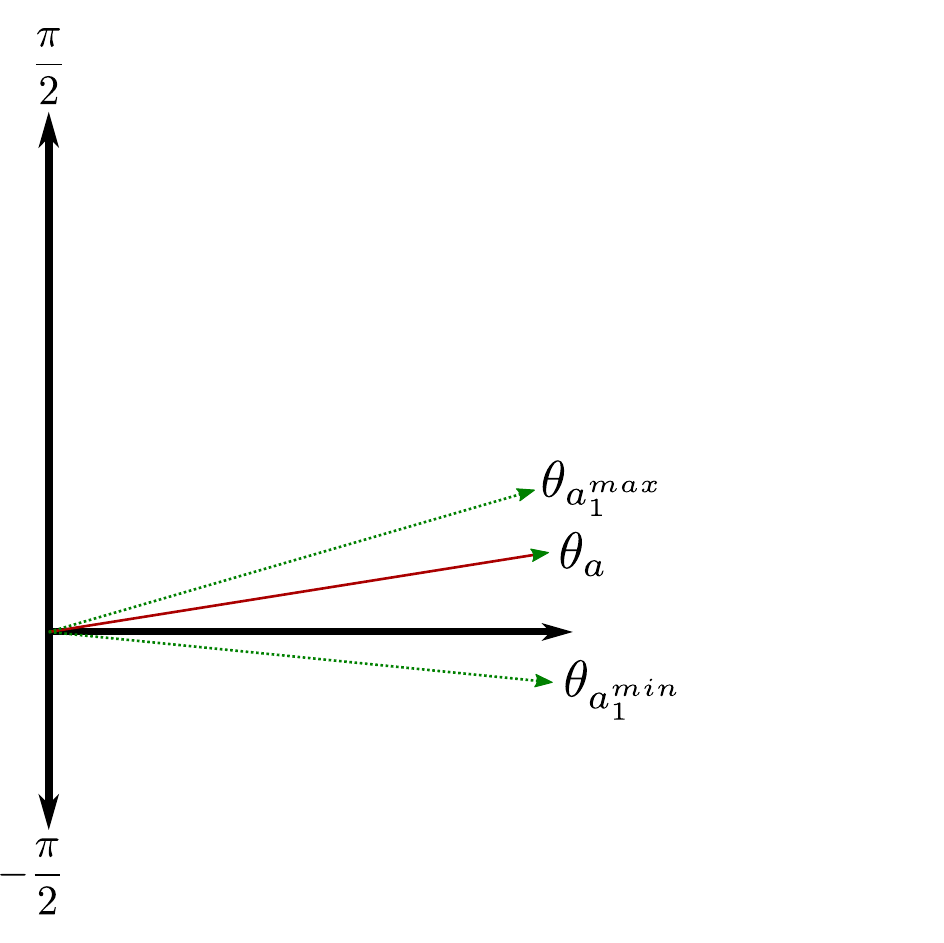}
        \label{fig:angle_problem_image}
    \end{subfigure}
    \caption{Left: the red dot corresponds to $a$, the probability to be estimated; $a_1^{\min}$ and $a_1^{\max}$ define the confidence interval whose width is $2\epsilon_1$. Right: the same, but represented in terms of the angles. Note that the ``true value" --represented by either $a$ and $\theta_a$-- falls in a generic position within the confidence interval.}
    \label{fig:angle_problem}
\end{figure}

From \eqref{ide} and \eqref{eqn:first_empirical_amplitude_estimation}, we can obtain a first confidence interval:
\begin{equation}\label{eq:first_amplitude_bounds}
      \begin{aligned}
    &a^{\max}_1 = \min\left(\dfrac{\hat{p}_{\text{sum}}-\hat{p}_{\text{diff}}}{4b_1}+\dfrac{\epsilon^{p}_1}{|2b_1|},1\right),\\
    &a^{\min}_1 = \max\left(\dfrac{\hat{p}_{\text{sum}}-\hat{p}_{\text{diff}}}{4b_1}-\dfrac{\epsilon^{p}_1}{|2b_1|},-1\right),\\
    &a_1 = \dfrac{a^{\max}_1+a^{\min}_1}{2}\ ,\\ 
    &\epsilon_1^a = \dfrac{a^{\max}_1-a^{\min}_1}{2}\ ,
\end{aligned} 
\end{equation}
where the $\max$ and the $\min$ operations are introduced because we know a priori that probabilities are bounded between $0$ and $1$. The assignment of an error $\epsilon^{p}_1$ to the empirical result $\hat{p}_1$ relies on a statistical analysis and depends on the statistical bound one employs, such as Chebysev, Chernoff (Hoeffding) and Clopper-Pearson bounds.
Here one of the main differences with respect to the other algorithms present in the literature becomes obvious, although the probabilities are bounded between $0$ and $1$ the estimated amplitude obtained by the identity \eqref{eqn:first_empirical_amplitude_estimation} is now bounded between $-1\leq a_1\leq 1$, that is, \textbf{it can be negative} (see Figure \ref{fig:angle_problem}). Note that the sign of the amplitude depends on the sign of $b_1$, which is taken as being positive for simplicity. However, this election is arbitrary and it could be chosen negative. 

\subsection{Following iterations: amplifying the probability and shrinking the interval}

\begin{figure}[H]
        \centering
    \begin{subfigure}[t]{0.3\textwidth}
        \centering
        \includegraphics[width=\textwidth]{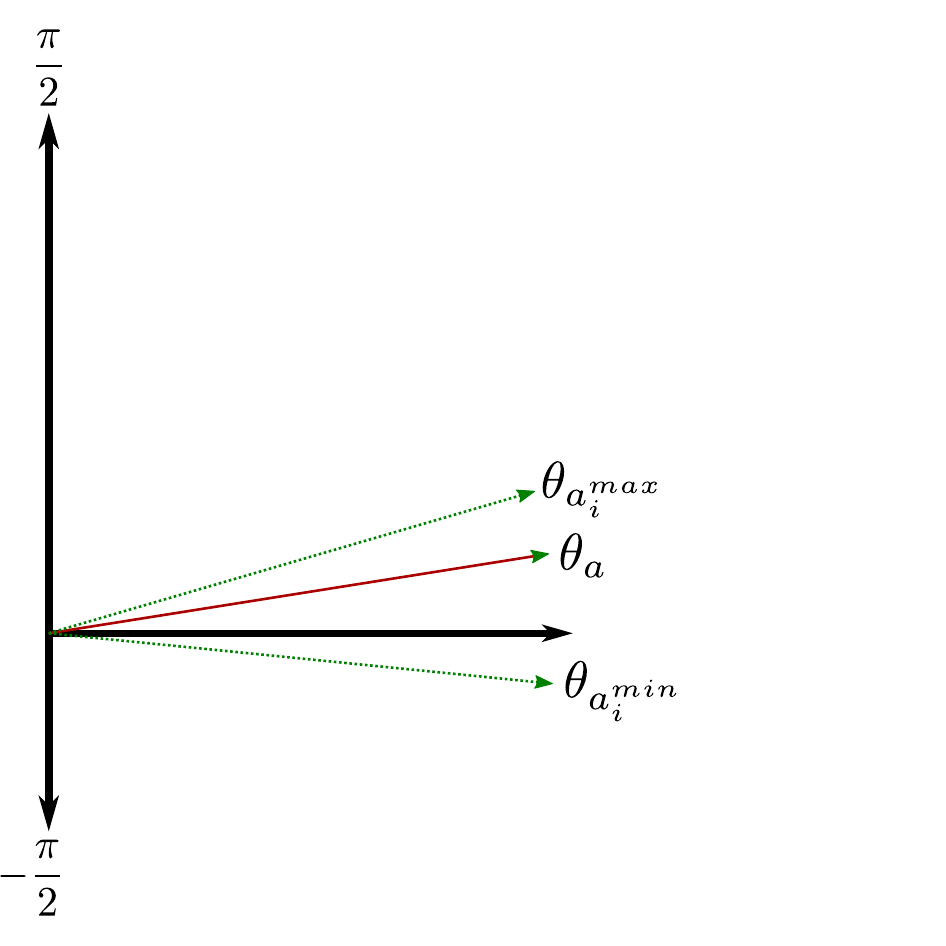}
        \caption{Starting point.}
        \label{fig:angle_problem_image2}
    \end{subfigure}
    \hfill
    \begin{subfigure}[t]{0.3\textwidth}
        \centering
        \includegraphics[width=\textwidth]{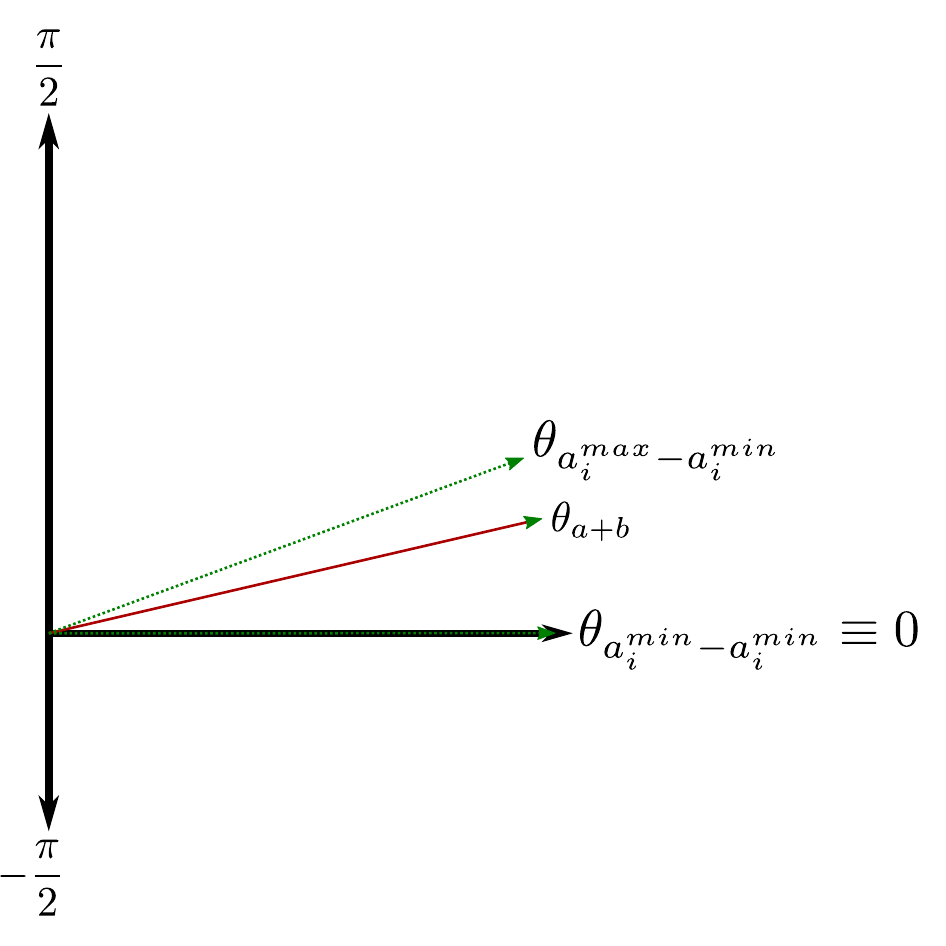}
        \caption{Shift.
        } \label{fig:shift}
    \end{subfigure}
    \hfill 
     \begin{subfigure}[t]{0.3\textwidth}
        \centering
        \includegraphics[width =\textwidth]{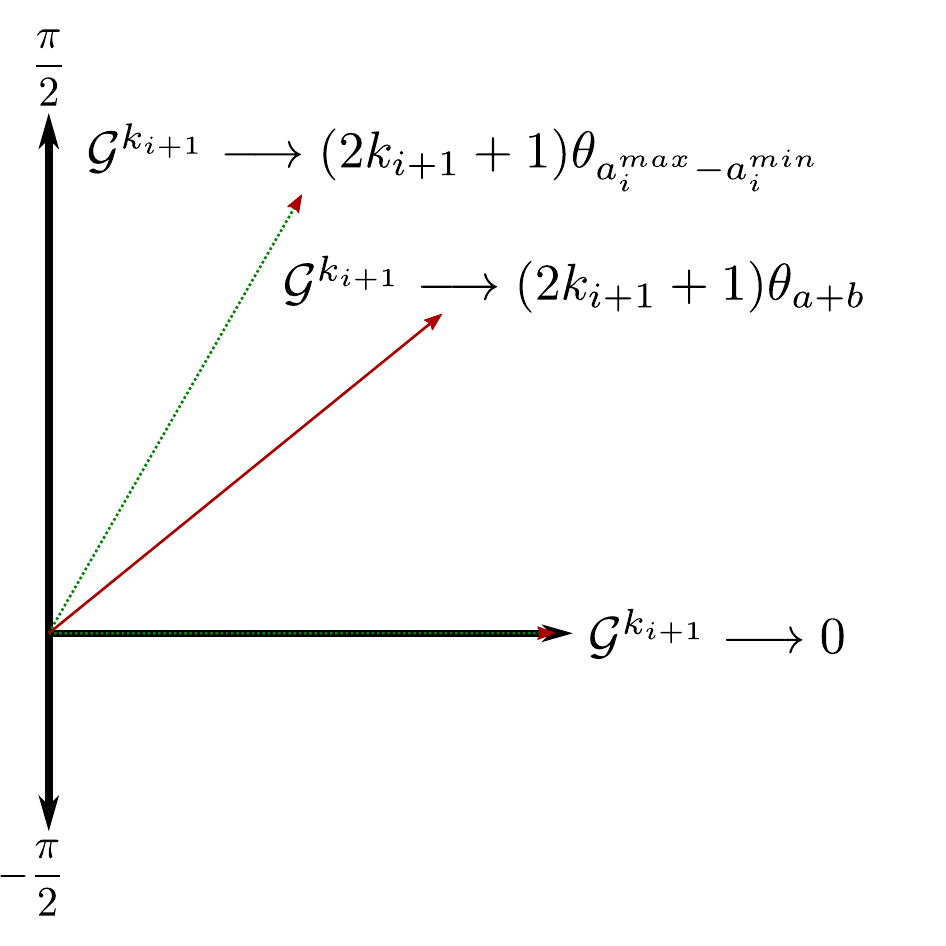}
        \subcaption{Amplification.
        }\label{fig:grover_operator}
    \end{subfigure}
        \\
     \begin{subfigure}[t]{0.3\textwidth}
        \centering
    \includegraphics[width=\textwidth]{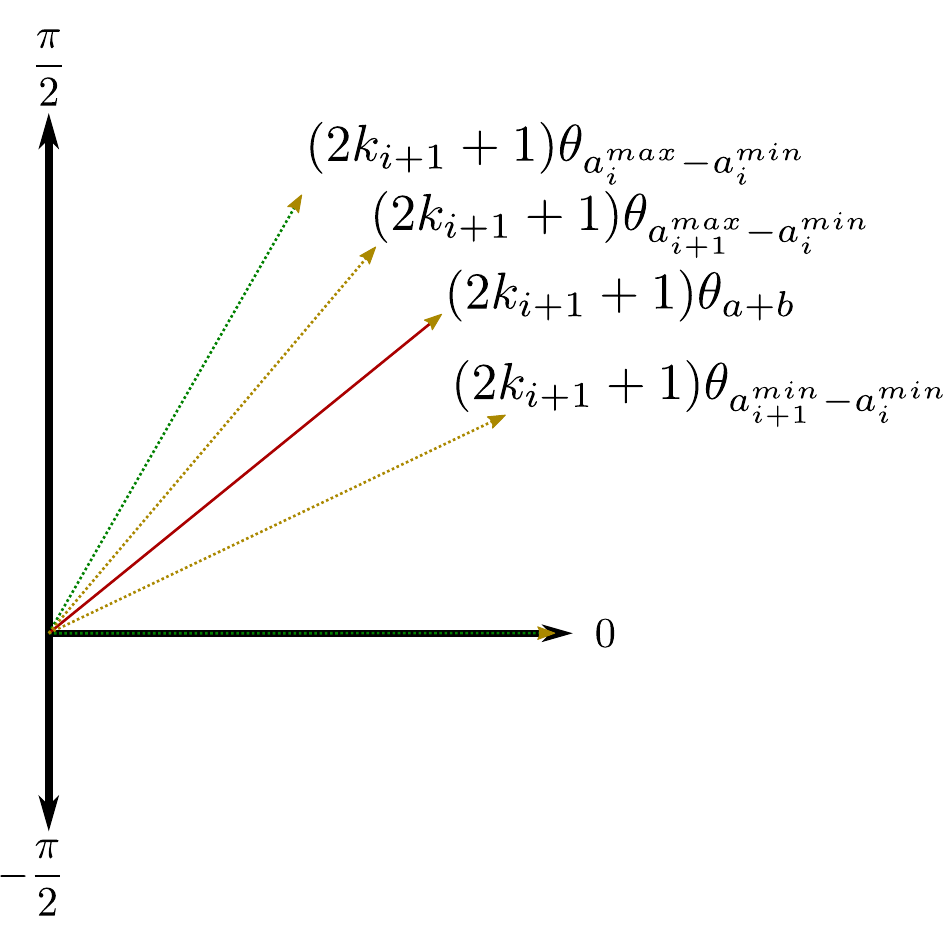}
    \subcaption{Measuring.
    }\label{fig:amplified_space_sampling}
    \end{subfigure}
\hfill
     \begin{subfigure}[t]{0.3\textwidth}
        \centering
        \includegraphics[width=\textwidth]{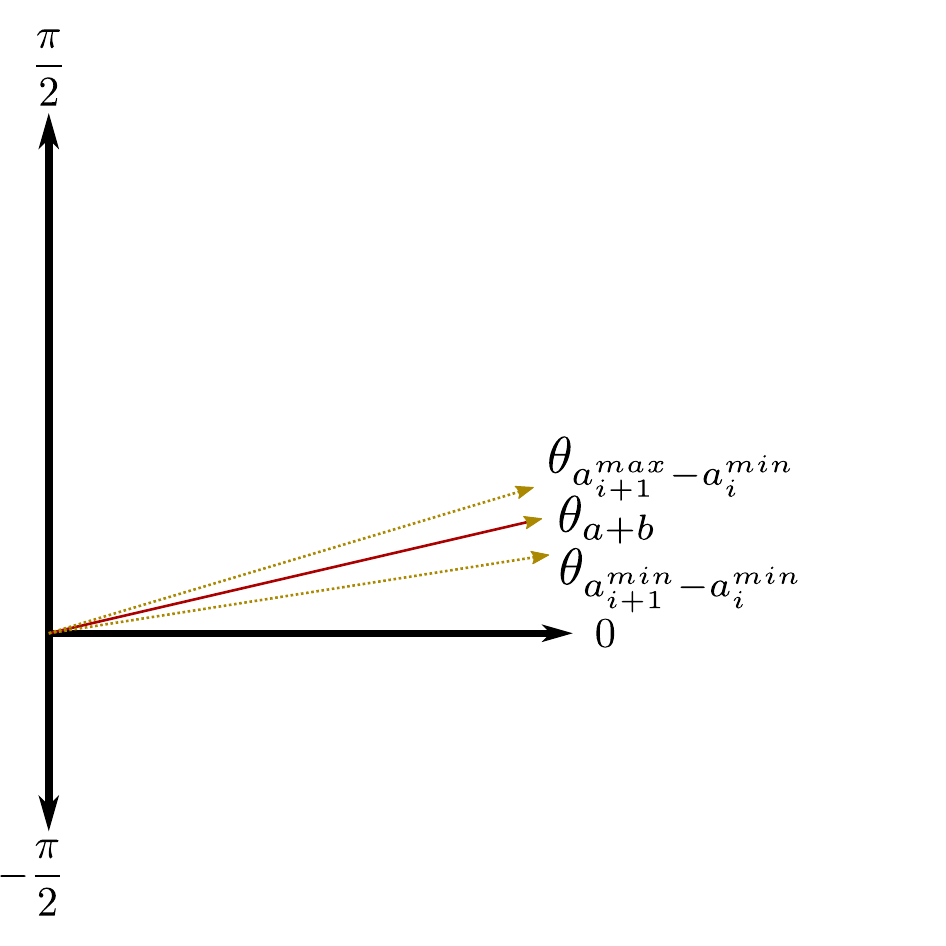}
         \subcaption{Undoing the amplification.
         }
    \label{fig:unamplified}
        \end{subfigure}
    \hfill
    \begin{subfigure}[t]{0.3\textwidth}
        \centering
        \includegraphics[width=\textwidth]{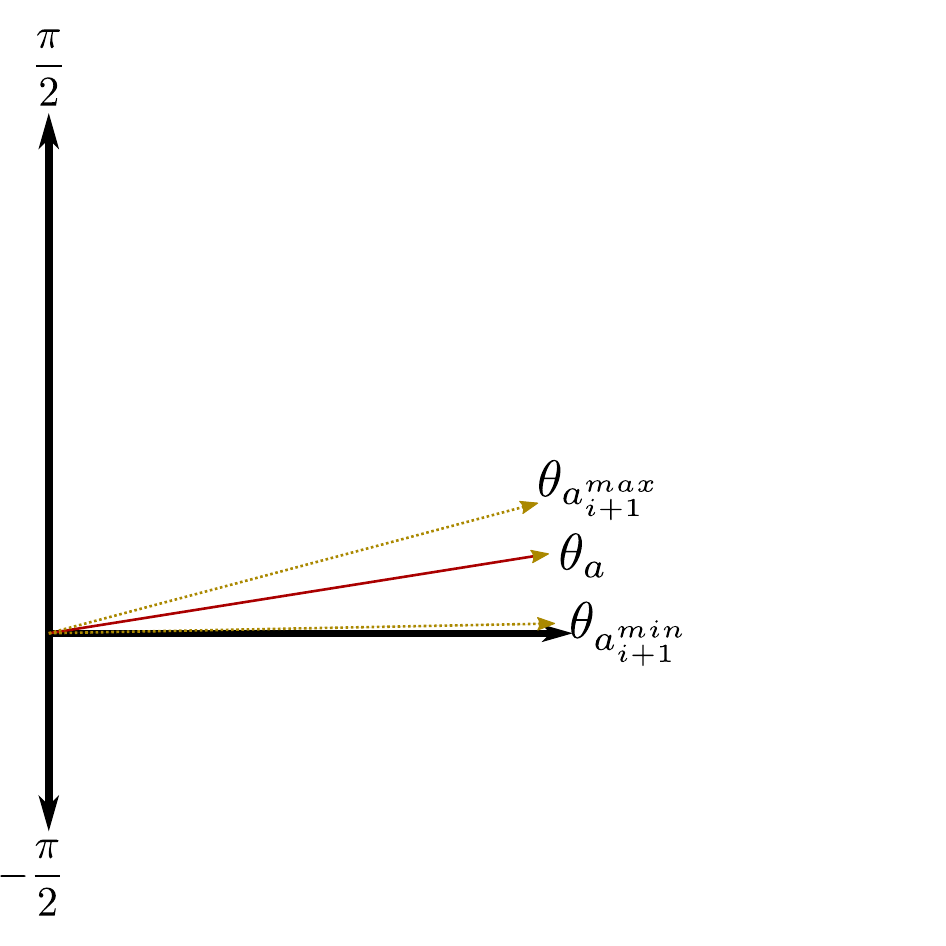}
        \subcaption{Undoing the shift.
        }\label{fig:second_iteration}
    \end{subfigure}
    \caption{Steps on each iteration.}
    \label{fig:following_iterations}
\end{figure}

On consecutive iterations, given an input confidence interval $(a^{\min}_i,a^{\max}_i)$ (see Figure \ref{fig:angle_problem_image2}) we want to obtain a tighter one $(a^{\max}_{i+1},a^{\min}_{i+1})$ and iterate the process until the desired precision $\epsilon$ is reached. At each iteration, the process for narrowing the interval starts by choosing a new shift according to 
\begin{equation}\label{shifti+1}
    b_{i+1} = -a^{\min}_{i}\ .
\end{equation}
This election is not unique. Again, we could have chosen  $b_{i+1} = -a^{\max}_{i}$. Always keep in mind that the phase that we are obtaining is relative to the original value of $b = b_1$.
Considering the choice \eqref{shifti+1}, we force our lower bound to match exactly zero (See Figure \ref{fig:shift}). The boundaries of the confidence interval $(a^{\min}_i,a^{\max}_i)$, when shifted and then expressed in terms of the corresponding angles, become:
\begin{equation}\label{alphamax}
\begin{aligned}
    &\alpha^{\max}_{i} = \arcsin\left(a^{\max}_i-a^{\min}_i\right) = \arcsin\left(2\epsilon^a_i \right),\\
    &\alpha^{\min}_{i} = 0.
\end{aligned}
\end{equation}
The angular region $\alpha_i^{\min} \leq \alpha_i \leq \alpha_i^{\max}$ represents the confidence interval and we refer to it as \emph{confidence fan}.\\ \\
The next step takes advantage of the Grover operator, defined as
\begin{equation}\label{groove}
    \mathcal{G} = -\mathcal{A}_{b}\mathcal{R}_{\ket{0}}\mathcal{A}_{b}^{\dagger}\mathcal{R}_{\ket{\phi}}\ ,
\end{equation}
where 
\begin{equation}
   \begin{aligned}
    \mathcal{R}_{\ket{0}} &=\mathbb{1}-2\ket{0}\bra{0}\ ,\\
    \mathcal{R}_{\ket{\phi}} &=\mathbb{1}-2\ket{\phi}\bra{\phi}\ ,
\end{aligned} 
\end{equation}
and $\mathcal{A}_{b}$ is the oracle defined in \eqref{eq:shi_def}.
The Grover operator applied $k_{i+1}$ times allows us to multiply the angles associated to the $i$-th bounds by $2k_{i+1}+1$ (see Figure \ref{fig:grover_operator}). When doing so, the distance measured in angles between the upper and the lower bounds increases by a factor $2k_{i+1}+1$:
\begin{align}
    &\ket{\psi_{i+1}}_+ = \left(a_i+b_{i+1}\right)\ket{\phi}+c_i\ket{\phi^\perp}= \left(a_i-a_i^{\min}\right)\ket{\phi}+c_i\ket{\phi^\perp} \equiv \sin(\theta_{i+1})\ket{\phi}+\cos(\theta_{i+1})\ket{\phi^\perp}\\ \nonumber
    &\qquad \xrightarrow{{\cal G}^{k_{i+1}}}\qquad 
    \sin\Big[(2k_{i+1}+1)\theta_{i+1}\Big]\ket{\phi}+\cos\Big[(2k_{i+1}+1)\theta_{i+1}\Big]\ket{\phi^\perp}\ ,
\end{align}
where ${\cal G}^{k_{i+1}}$ indicates the Grover operator applied $k_{i+1}$ times. In order to avoid ambiguities due to the lack of a bijective correspondence angle/amplitude, when measuring amplified probabilities, we cannot allow the amplified angles to go beyond $\left[0,\dfrac{\pi}{2}\right]$. Namely, we need the amplified confidence fan to stay within the first quadrant. Relying on \eqref{alphamax}, we choose the Grover amplification exponent as:
\begin{equation}\label{eq:next_k}
k_{i+1} = \left\lfloor\dfrac{\pi}{4\arcsin\left(2\epsilon^a_{i}\right)}-\dfrac{1}{2}\right\rfloor\ ,
\end{equation}
so that we maximize the amplification factor while respecting the angle constraint.\\ \\

Now, we measure the state $\ket{\psi_{i+1}}$ in the amplified space, obtaining the empirical probability 
\begin{equation}
    \hat{p}_{i+1}\approx \sin^2 \Big[(2k_{i+1}+1)\theta_{i+1}\Big]\ ,
\end{equation} 
with the statistical error $ \epsilon^{p}_{i+1}$, and define:
\begin{align}\nonumber
    & p^{\max}_{i+1} := \min\left(\hat{p}_{i+1}+\epsilon^{p}_{i+1},1\right)\ ,\\ \label{pmaxmin}
    & p^{\min}_{i+1} := \max\left(\hat{p}_{i+1}-\epsilon^{p}_{i+1},0\right)\ ,\\ \nonumber
    & p_{i+1} := \dfrac{p^{\max}_{i+1}+p^{\min}_{i+1}}{2}\ ,
\end{align}
where the $\max$ and $\min$ functions play an analogous role as in Section \ref{sign} (see Figure \ref{fig:amplified_space_sampling}).\\ \\
In the next step we transform the angles corresponding to $p^{\max}_{i+1}$ and $p^{\min}_{i+1}$ to the non-amplified space:
\begin{equation}
    \begin{aligned}
    & \theta^{\max}_{i+1} = \dfrac{\arcsin\left(\sqrt{p^{\max}_{i+1}}\right)}{2k_{i+1}+1}\ ,\\
    & \theta^{\min}_{i+1} =  \dfrac{\arcsin\left(\sqrt{p^{\min}_{i+1}}\right)}{2k_{i+1}+1}\ .
\end{aligned}
\end{equation}
In other words, we have just ``undone" the amplification (see Figure \ref{fig:unamplified}). \\ \\

Finally, we have to undo the shift \eqref{shifti+1}, actually performing an opposite shift (see Figure \ref{fig:second_iteration}). Using definitions analogous to those given in \eqref{eq:first_amplitude_bounds}, we
finally obtain:
\begin{equation}
   \begin{aligned}\label{eqn:following_amplitude_bounds}
    & a^{\max}_{i+1} = \sin\left(\dfrac{\arcsin\left(\sqrt{p^{\max}_{i+1}}\right)}{2k_{i+1}+1}\right)-b_{i+1}\ ,\\
    & a^{\min}_{i+1} =  \sin\left(\dfrac{\arcsin\left(\sqrt{p^{\min}_{i+1}}\right)}{2k_{i+1}+1}\right)-b_{i+1}\ ,\\
    & a_{i+1} = \dfrac{a^{\max}_{i+1}+a^{\min}_{i+1}}{2}\ ,\\
    & \epsilon^a_{i+1} = \dfrac{a^{\max}_{i+1}-a^{\min}_{i+1}}{2} = \frac{1}{2}\sin\left(\dfrac{\arcsin\left(\sqrt{p^{\max}_{i+1}}\right)}{2k_{i+1}+1}\right)-\frac{1}{2}\sin\left(\dfrac{\arcsin\left(\sqrt{p^{\min}_{i+1}}\right)}{2k_{i+1}+1}\right).
\end{aligned} 
\end{equation}

Recall that the goal is to reduce the width of the confidence interval until the desired precision $\epsilon$ is reached. To this sake, one has to repeat the iteration just described until the goal is met.\\ \\


\begin{remark}
Throughout this section we have addressed the general structure of the algorithm. Nevertheless, we have not specified the values of all the parameters involved. More specifically, we have not discussed how $\epsilon^p_i$ is obtained. This parameter strictly depends on the number of shots on each iteration, $N_i$, and the confidence required on each iteration, $1-\gamma_i$, through a set of bounds such as Hoeffding's inequality or Clopper-Pearson bound (see \cite{hoeffding,clopper_pearson}). In Section \ref{sec:configuration_properties}, more insight about these choices is provided.
\end{remark}

\section{RQAE: configuration and properties}\label{sec:configuration_properties}
    As mentioned before, in order to complete the RQAE method, we need to incorporate a particular choice for the parameter $\epsilon^p_i$ and, thus, the parameters involved in its computation, \emph{i.e.}, $N_i$ and $1-\gamma_i$, the number of shots and confidence level of the $i$-th iteration, respectively. With the aim of being able to characterize theoretically the algorithm, we propose to take, $\forall i$, the following constant values:
    \begin{equation*}
    \begin{aligned}
        N_i(q,\epsilon,\gamma) &= \left\lceil \dfrac{1}{2\epsilon^p(q)^2}\log\left(\dfrac{2T(q,\epsilon)}{\gamma}\right) \right\rceil, \\
        \gamma_i(q,\epsilon,\gamma) &= \dfrac{\gamma}{T(q,\epsilon)},\\
        b_1(q) &= \dfrac{1}{2}\sin\left(\dfrac{\pi}{2(q+2)}\right),
    \end{aligned}
    \end{equation*}
    where
    \begin{equation*}
        \epsilon^p (q) = \dfrac{1}{2}\sin^2\left(\dfrac{\pi}{2(q+2)}\right) \quad \text{and} \quad T(q,\epsilon) = \log_q\left(q^2\dfrac{\arcsin\left(\sqrt{2\epsilon^p(q)}\right)}{\arcsin\left(2\epsilon\right)}\right).
    \end{equation*}
    Once the free parameters are selected, the RQAE is completed and ready to be used. Of course, other choices can be made. In Algorithm \ref{alg:RQAE}, the RQAE algorithm is schematically described. Note that our selection of the parameters is uniquely determined by the input quantities $q$, $\epsilon$ and $\gamma$. As we will see in Section \ref{sec:properties}, the choice considered here presents several interesting properties.

\begin{algorithm}[H]
\caption{RQAE pseudocode}\label{alg:RQAE}
\begin{algorithmic}
\State \textbf{Input:}
\Indent
\State $\epsilon$ \Comment{required precision}
\State $\gamma$ \Comment{$1-\gamma$ is the confidence level}
\State $q$  \Comment{amplification policy}
\State $\mathcal{A}$ \Comment{oracle}
\EndIndent
\State \textbf{Output:}
\Indent
\State $a$ \Comment{estimated amplitude with sign}
\EndIndent
\State \textbf{Algorithm:}
\Indent
\State \Comment{Define relevant parameters}
\State Set $\epsilon^p = \dfrac{1}{2}\sin^2\left(\dfrac{\pi}{2(q+2)}\right)$
\State Set $T = \log_q\left(q^2\dfrac{\arcsin\left(\sqrt{2\epsilon^p}\right)}{\arcsin\left(2\epsilon\right)}\right)$
\State Set $\gamma_i = \dfrac{\gamma}{T}$\Comment{Confidence for each iteration}
\State Set $N_i = \left\lceil \dfrac{1}{2(\epsilon^p)^2}\log\left(\dfrac{2T}{\gamma}\right) \right\rceil$ \Comment{Number of shots for each iteration}
\State Set $\epsilon^p_i = \sqrt{\dfrac{1}{2N_i}\log\left(\dfrac{2}{\gamma_i}\right)}$
\State Set $k^{\max}=\left\lceil \dfrac{1}{2}\dfrac{\arcsin\left(\sqrt{2\epsilon^p}\right)}{\arcsin\left(2\epsilon\right)}-\dfrac{1}{2}\right\rceil$
\State \Comment{First Iteration}
\State Set $b = \dfrac{1}{2}\sin\left(\dfrac{\pi}{2(q+2)}\right)$\Comment{Shift}
\State Measure $p_{\text{sum}}$ and $p_{\text{diff}}$
\State $a^{\max} = \min\left(\dfrac{\hat{p}_{\text{sum}}-\hat{p}_{\text{diff}}}{4b}+\dfrac{\epsilon^{p}_i}{|2b|},1\right)$
\State $a^{\min} = \max\left(\dfrac{\hat{p}_{\text{sum}}-\hat{p}_{\text{diff}}}{4b}-\dfrac{\epsilon^{p}_i}{|2b|},-1\right)$
\State $a = \dfrac{a^{\max}+a^{\min}}{2}$
\State $\epsilon^a = \dfrac{a^{\max}-a^{\min}}{2}$
\State \Comment{Following Iterations}
\While{$\epsilon^a>\epsilon$}
\State Set $b = -a^{\min}$ \Comment{Shift}
\State Set $k = \left\lfloor\dfrac{\pi}{4\arcsin(2\epsilon^a)}-\dfrac{1}{2} \right\rfloor$ \Comment{Number of amplifications}
\If{$k>k^{\max}$} 
\State $k = k^{\max}$
\EndIf
\State Measure $p$ \Comment{shifted probability with $k$ amplifications}
\State $p^{\max}= \min(p+\epsilon^p_i,1)$
\State $p^{\min}= \max(p-\epsilon^p_i,0)$
\State $\theta^{\max} = \dfrac{\arcsin(\sqrt{p^{\max}})}{2k+1}$
\State $\theta^{\min} = \dfrac{\arcsin(\sqrt{p^{\min}})}{2k+1}$
\State $a^{\max} = \sin\left(\theta^{\max}\right)-b $
\State $ a^{\min} =  \sin\left(\theta^{\min}\right)-b$
\State $a = \dfrac{a^{\max}+a^{\min}}{2}$
\State $\epsilon^a = \dfrac{a^{\max}-a^{\min}}{2}$
\EndWhile
\State \textbf{return} $a$
\EndIndent
\end{algorithmic}
\end{algorithm}

\subsection{Properties}\label{sec:properties}
Given the proposed configuration in Algorithm \ref{alg:RQAE}, the RQAE algorithm presents several properties which are listed in the next theorem.
\begin{theorem}\label{the:properties}

Given $\epsilon,\gamma$ and $q$, and taking the parameters:
\begin{equation}\label{eq:number_shots}
     N_i(q,\epsilon,\gamma) =\left\lceil N \right\rceil = \left\lceil \dfrac{1}{2\epsilon^p(q)^2}\log\left(\dfrac{2T(q,\epsilon)}{\gamma}\right) \right\rceil,
\end{equation}
\begin{equation}\label{eq:max_confidence}
    \gamma_i(q,\epsilon,\gamma) = \dfrac{\gamma}{T(q,\epsilon)},
\end{equation}
\begin{equation}\label{eq:b1}
    b_1 = \dfrac{1}{2}\sin\left(\dfrac{\pi}{2(q+2)}\right)\ ,
\end{equation}
with
\begin{equation}\label{eq:min_precision}
    \epsilon^p (q) = \dfrac{1}{2}\sin^2\left(\dfrac{\pi}{2(q+2)}\right)\ ,
\end{equation}
\begin{equation}\label{eq:max_iterations}
    T(q,\epsilon) = \log_q\left(q^2\dfrac{\arcsin\left(\sqrt{2\epsilon^p(q)}\right)}{\arcsin\left(2\epsilon\right)}\right)\ ,
\end{equation}
then
\begin{enumerate}
    \item The error at each iteration is bounded by:
    \begin{equation}\label{eq:precision_bound}
        \epsilon^p_i\leq \epsilon^p
    \end{equation}
    \item We get the amplification policy:
    \begin{equation}\label{eq:amplification_policy_bound}
        q_i = \dfrac{2k_{i+1}+1}{2k_i+1}\geq q
    \end{equation}
    \item The depth of the circuit is bounded by:
    \begin{equation}\label{eq:depth_bound}
        k_I\leq \left\lceil\dfrac{1}{2}\dfrac{\arcsin\left(\sqrt{2\epsilon^p}\right)}{\arcsin(2\epsilon)}-\dfrac{1}{2}\right\rceil =k_I^{\max}
    \end{equation}
    \item The algorithm finishes before $T(q,\epsilon)$ iterations:
    \begin{equation}\label{eq:max_iterations_bound}
        T>I
    \end{equation}
    \item The algorithm obtains a precision $\epsilon$ with confidence $1-\gamma$ (Proof of Correctness):
    \begin{equation}\label{eq:master_equation}
        \mathbb{P}\Big[a\not\in (a^{\min}_I,a^{\max}_I)\Big]\leq \gamma
    \end{equation}
    
    \item The total number of calls to the oracle is bounded by:
    \begin{align}\label{eq:oracle_bound}
    N_{\text{oracle}}<&\dfrac{1}{\sin^4\left(\dfrac{\pi}{2(q+2)}\right)}\log\left[\dfrac{2\sqrt{e}\log_q\left(\dfrac{q^2\pi}{2(q+2)\arcsin(2\epsilon)}\right)}{\gamma}\right]\\ \nonumber &\qquad\cdot\left(\dfrac{\pi}{2(q+2)\arcsin(2\epsilon)}+2\right)\left(1+\dfrac{q}{q-1}\right)\ .
     \end{align} 
\end{enumerate}
\end{theorem}

\noindent
The proof of Theorem \ref{the:properties} can be found in Appendix \ref{sec:proof}.\\

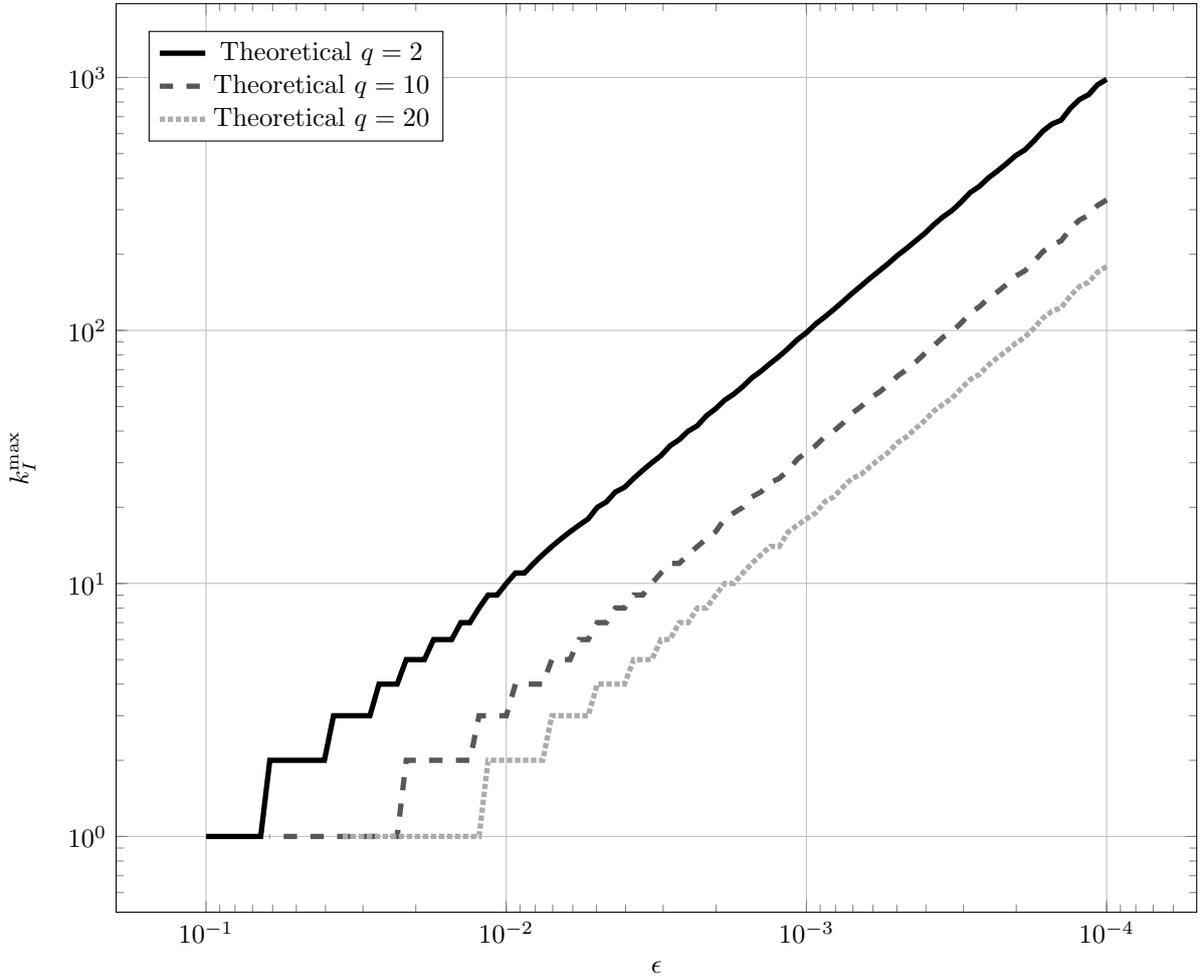
\begin{figure}[H]
    \centering
    \begin{tikzpicture}
    \begin{axis}[
    xmajorgrids=true,
    ymajorgrids=true,
    width=\textwidth,
    legend pos=north west,
    xmode =log,
    ymode=log,
    x dir=reverse,
    xlabel=$\epsilon$,
    ylabel=$k_I^{\max}$,
    ]
\addplot [
    domain=0.0001:0.1, 
    samples=100, 
    color=black,
    line width = 2pt,
    style = solid,
]
{ceil(0.5*pi*180/(2*(2+2)*pi*asin(2*x))-0.5)};
\addlegendentry[]{Theoretical $q = 2$}
\addplot [
    domain=0.0001:0.1, 
    samples=100, 
    color=black!66!white,
    line width = 2pt,
    dash pattern=on 5pt off 5pt,
]
{ceil(0.5*pi*180/(2*(10+2)*pi*asin(2*x))-0.5)};
\addlegendentry[]{Theoretical $q = 10$}
\addplot [
    domain=0.0001:0.1, 
    samples=100, 
    color=black!33!white,
    line width = 2pt,
    style = densely dotted,
]
{ceil(0.5*pi*180/(2*(20+2)*pi*asin(2*x))-0.5)};
\addlegendentry[]{Theoretical $q = 20$}
\end{axis}
\end{tikzpicture}
\caption{Theoretical maximum depth of the circuit $k_I^{\max}$ in terms of the precision $\epsilon$. }
\label{fig:depth}
\end{figure}

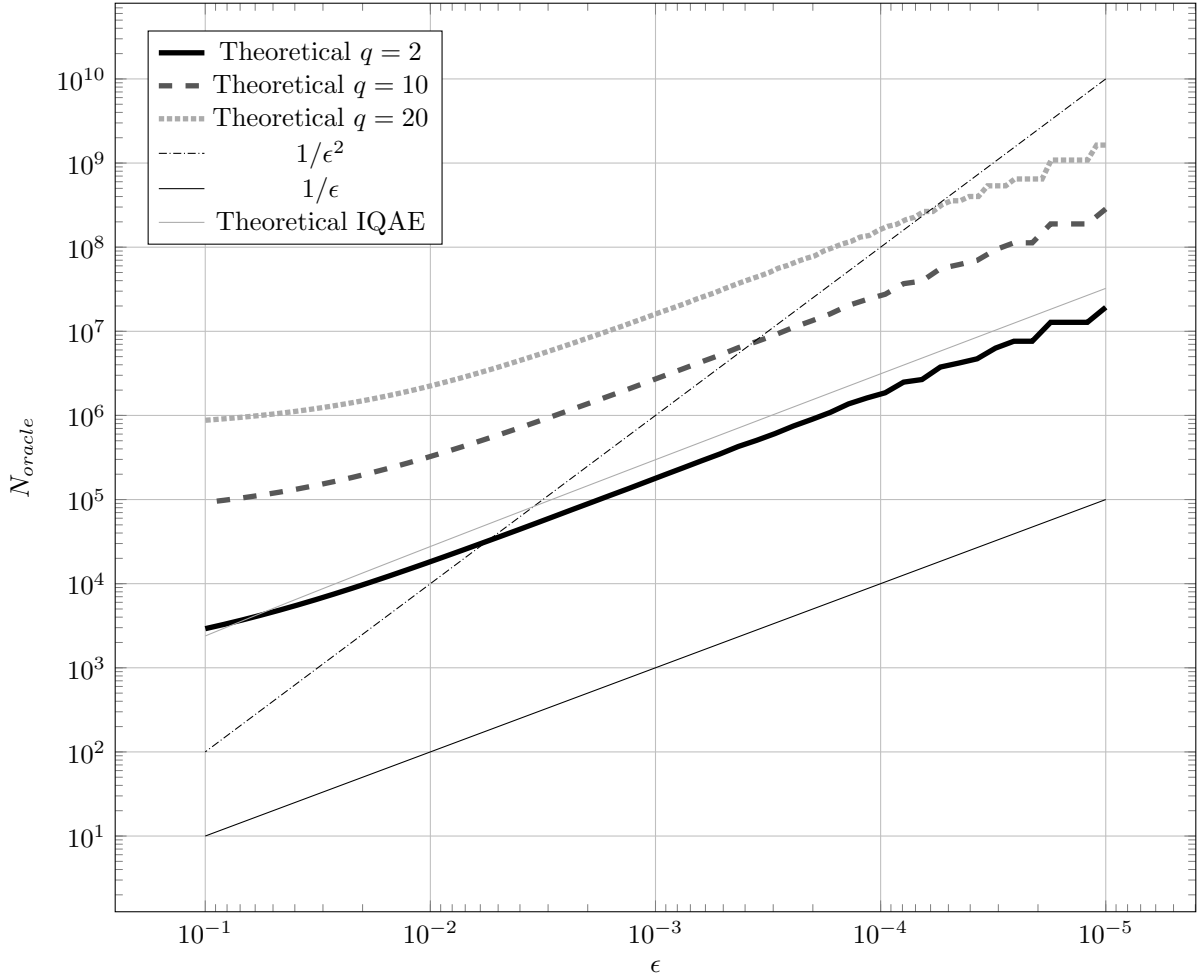
\begin{figure}[H]
    \centering
    \begin{tikzpicture}
    \begin{axis}[
    xmajorgrids=true,
    ymajorgrids=true,
    width=\textwidth,
    legend pos=north west,
    xmode =log,
    ymode=log,
    x dir=reverse,
    xlabel=$\epsilon$,
    ylabel=$N_{oracle}$,
    ]
\addplot [
    domain=0.00001:0.1, 
    samples=50, 
    color=black,
    line width = 2pt,
    style = solid,
]
{ln((2*sqrt(e)*ln(2*2*pi*180/(2*(2+2)*pi*asin(2*x))))/(0.05*ln(2)))*1/(sin(deg(pi/(2*(2+2))))^4)*(pi*180/(2*(2+2)*pi*asin(2*x))+2)*(1+2/(2-1))};
\addlegendentry[]{Theoretical $q = 2$}
\addplot [
    domain=0.00001:0.1, 
    samples=50, 
    color=black!66!white,
    line width = 2pt,
    dash pattern=on 5pt off 5pt,
]
{ln((2*sqrt(e)*ln(10*10*pi*180/(2*(10+2)*pi*asin(2*x))))/(0.05*ln(10)))*1/(sin(deg(pi/(2*(10+2))))^4)*(pi*180/(2*(10+2)*pi*asin(2*x))+2)*(1+10/(10-1))};
\addlegendentry[]{Theoretical $q = 10$}
\addplot [
    domain=0.00001:0.1, 
    samples=100, 
    color=black!33!white,
    line width = 2pt,
    style = densely dotted,
]
{ln((2*sqrt(e)*ln(20*20*pi*180/(2*(20+2)*pi*asin(2*x))))/(0.05*ln(20)))*1/(sin(deg(pi/(2*(20+2))))^4)*(pi*180/(2*(20+2)*pi*asin(2*x))+2)*(1+20/(20-1))};
\addlegendentry[]{Theoretical $q = 20$}
\addplot [
domain=0.00001:0.1, 
samples=100, 
color=black,
style=densely dashdotted,
]
{1/x^2};
\addlegendentry[]{$1/\epsilon^2$}
\addplot [
    domain=0.00001:0.1, 
    samples=100, 
    color=black,
]
{1/x};
\addlegendentry{$1/\epsilon$}
\addplot [
    domain=0.00001:0.1, 
    samples=100, 
    color=black!33!white,
]
{50/x*ln(2/0.05*ln(pi/(4*x))/ln(2))};
\addlegendentry{Theoretical IQAE}
\end{axis}
\end{tikzpicture}
\caption{Theoretical maximum number of calls to the oracle $N_{\text{oracle}}$ in terms of the precision $\epsilon$ for different values of the amplification policy $q$ and a fixed confidence level $1-\gamma = 0.95$. Lines $1/\epsilon$ and $1/\epsilon^2$ 
are depicted for comparison purposes.}
\label{fig:theoretical_calls}
\end{figure}

The first important feature is the fourth property, where we see that the algorithm achieves the desired precision $\epsilon$ with confidence at least $1-\gamma$, which was our goal in the beginning (see Equation \eqref{eq:initial_problem}).
Property number two justifies why we call the parameter $q$ the ``amplification policy''. The reason for this is that $q$ is indeed the minimum ratio of the amplification of an iteration with respect to the previous, thus controlling the amplification policy.
The depth of the circuit is intimately related with the amplification policy. In property number three we have a clear bound $k_I^{\max}$ for the depth. To get a clearer idea of this bound, in Figure \ref{fig:depth} we depict the maximum depth of the circuit $k_I^{\max}$ in terms of the precision $\epsilon$ for different amplification policies $q$.
As we see, the depth decreases with $q$, but at the cost of increasing the number of shots on each iteration (see Equation \eqref{eq:number_shots}). This directly leads to the question about what is more relevant in relation to the total number of shots: either performing more shots for each iteration with less iterations (and thus less circuit depth) or performing less shots at the cost of increasing the total number of iterations. 
This question is directly answered with the sixth property, where we have a bound for the number of shots in terms of the required precision $\epsilon$, the confidence $1-\gamma$ and the amplification policy $q$. To facilitate the interpretation of the expression, in Figure \ref{fig:theoretical_calls} we represent the theoretical number of calls to the oracle in terms of the precision $\epsilon$ for different amplification policies $q$ and a fixed confidence level of $1-\gamma = 0.95$ .
The dash dotted line, the black solid thin line and the gray solid thin lines are there for comparison purposes. We know that, without amplification, the number of executions of the circuit needed to achieve a precision $\epsilon$ grows as:
\begin{equation*}
    N_{oracle}\sim \dfrac{1}{\epsilon^2},
\end{equation*}
as it is given by classical bounds such as Chebysev or Clopper-Pearson. A quadratic speedup would then be obtaining the same precision with a total number of oracle calls:
\begin{equation*}
    N_{oracle}\sim \dfrac{1}{\epsilon}.
\end{equation*}
As we see in Figure \ref{fig:theoretical_calls} the number of calls needed in our method to achieve a precision $\epsilon$ grows approximately as $1/\epsilon$. This means that we have approximately achieved a quadratic speedup compared with unamplified  sampling.
Furthermore, we see that it is more efficient in terms of the total number of calls to the oracle to use lower amplification policies.

In general, the appropriate amplification policy depends on each specific case. On the one hand, the optimal $q$ depends both on the precision $\epsilon$ and the confidence level $1-\gamma$. On the other hand, in real hardware it could be more interesting to choose higher values of $q$, as we can not run arbitrarily long circuits.

Last, we have depicted the theoretical bound for the IQAE algorithm which, to our knowledge, represents the state of the art. As we see, the RQAE performance is better than that of the IQAE for some values of $q$. Note that the comparison with IQAE is not direct since they are estimating the probability and we are estimating the amplitude. However, in Appendix \ref{sec:probability_vs_amplitude} we show that, under some weak assumptions, this bounds are comparable.
\subsection{Empirical performance}\label{sec:experiments}

The sixth property of Theorem \ref{the:properties} gives us an upper bound to the number of calls to the oracle. However, it is always interesting to see how this theoretical bound compares with a real execution of the algorithm. Using the Quantum Learning Machine (QLM) developed by Atos we build a circuit with a total of $5$ qubits and we estimate one of the amplitudes encoded in the circuit with our method. Our algorithm is executed with a confidence level of $1-\gamma = 0.95$ and different amplification policies $q \in \{2,10,20\}$. For each level of target precision $\epsilon$, we perform $100$ experiments.
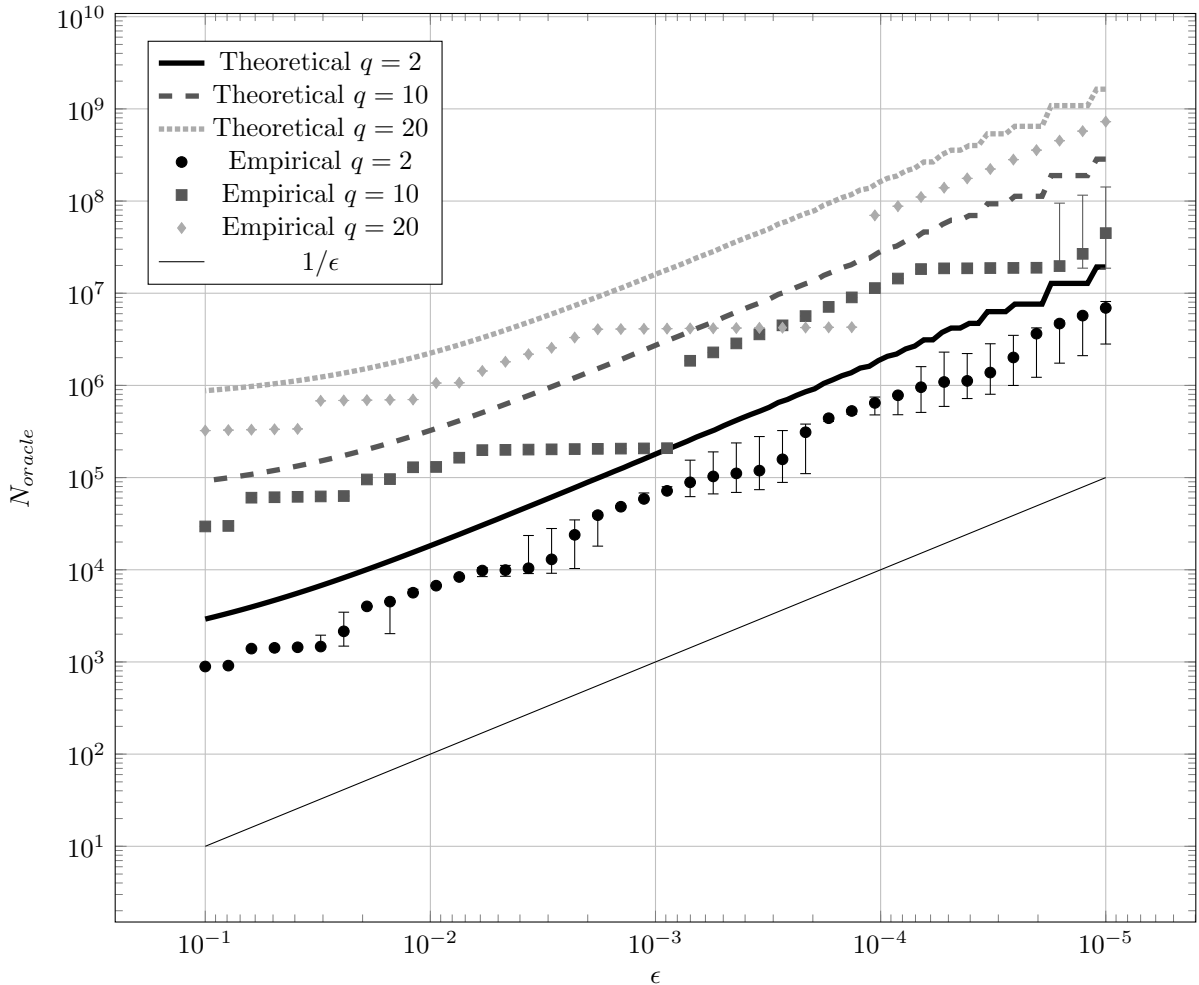
\begin{figure}[H]
    \centering
    \begin{tikzpicture}
    \begin{axis}[
    xmajorgrids=true,
    ymajorgrids=true,
    width=\textwidth,
    legend pos=north west,
    xmode =log,
    ymode=log,
    x dir=reverse,
    xlabel=$\epsilon$,
    ylabel=$N_{oracle}$,
    ]
\addplot [
    domain=0.00001:0.1, 
    samples=100, 
    color=black,
    line width = 2pt,
]
{ln((2*sqrt(e)*ln(2*2*pi*180/(2*(2+2)*pi*asin(2*x))))/(0.05*ln(2)))*1/(sin(deg(pi/(2*(2+2))))^4)*(pi*180/(2*(2+2)*pi*asin(2*x))+2)*(1+2/(2-1))};
\addlegendentry[]{Theoretical $q = 2$}
\addplot [
    domain=0.00001:0.1, 
    samples=100, 
    color=black!66!white,
    line width = 2pt,
    dash pattern=on 5pt off 5pt,
]
{ln((2*sqrt(e)*ln(10*10*pi*180/(2*(10+2)*pi*asin(2*x))))/(0.05*ln(10)))*1/(sin(deg(pi/(2*(10+2))))^4)*(pi*180/(2*(10+2)*pi*asin(2*x))+2)*(1+10/(10-1))};
\addlegendentry[]{Theoretical $q = 10$}
\addplot [
    domain=0.00001:0.1, 
    samples=100, 
    color=black!33!white,
    line width = 2pt,
    style=densely dotted,
]
{ln((2*sqrt(e)*ln(20*20*pi*180/(2*(20+2)*pi*asin(2*x))))/(0.05*ln(20)))*1/(sin(deg(pi/(2*(20+2))))^4)*(pi*180/(2*(20+2)*pi*asin(2*x))+2)*(1+20/(20-1))};
\addlegendentry[]{Theoretical $q = 20$}
\addplot[
    only marks,
    color=black,
    mark=*,
    error bars/.cd,
    y dir=both,
    y explicit,
    ]%
table[
    x=x,
    y=y,
    y error plus=y-max,
    y error minus=y-min,
    ]{data/q2_oracle_calls.dat};
\addlegendentry[]{Empirical $q = 2$}
\addplot[
    only marks,
    color=black!66!white,
    mark=square*,
    error bars/.cd,
    y dir=both,
    y explicit,
    ]%
table[
    x=x,
    y=y,
    y error plus=y-max,
    y error minus=y-min,
    ]{data/q10_oracle_calls.dat};
\addlegendentry[]{Empirical $q = 10$}
\addplot[
    only marks,
    color=black!33!white,
    mark=diamond*,
    error bars/.cd,
    y dir=both,
    y explicit,
    ]%
table[
    x=x,
    y=y,
    y error plus=y-max,
    y error minus=y-min,
    ]{data/q20_oracle_calls.dat};
\addlegendentry[]{Empirical $q = 20$}
\addplot [
    domain=0.00001:0.1, 
    samples=100, 
    color=black,
]
{1/x};
\addlegendentry{$1/\epsilon$}
\end{axis}
\end{tikzpicture}
\caption{Number of calls to the oracle, $N_{\text{oracle}}$, versus the target precision $\epsilon$. The plot is in a log-log scale with the $x$-axis inverted. The experimental points (circles, squares and diamonds) represent the mean number of oracle calls obtained over 100 experiments each, while the error bars stretch from the maximum to the minimum value obtained in the same set of experiments. The lines are the corresponding theoretical bounds from Equation \eqref{eq:oracle_bound}.}
\label{fig:experimental_calls}
\end{figure}
As we see in Figure \ref{fig:experimental_calls}, the experiments respect the theoretical bound for the total number of calls to the oracle. In particular, the theoretical bound proves not to be loose, meaning that --in general-- the experimental results keep close to the theoretical bound. Furthermore, we underline that the empirical behavior of $N_{\text{oracle}}$ with respect to $\epsilon$ follows the qualitative trend expected theoretically.

\begin{figure}[H]
    \centering
\begin{tikzpicture}
    \begin{axis}[
    xmajorgrids=true,
    ymajorgrids=true,
    width=0.49\textwidth,
    legend pos=north west,
    legend style={nodes={scale=0.85, transform shape}},
    xmode =log,
    ymode=log,
    x dir=reverse,
    xlabel=$\epsilon$,
    ylabel=$k_I$,
    ]
\addplot [
    domain=0.00001:0.1, 
    samples=100, 
    color=black,
    line width = 2pt,
    style = solid,
]
{ceil(0.5*pi*180/(2*(2+2)*pi*asin(2*x))-0.5)};
\addlegendentry[]{Theoretical $q = 2$}
\addplot [
    domain=0.00001:0.1, 
    samples=100, 
    color=black!33!white,
    line width = 2pt,
    style = densely dotted,
]
{ceil(0.5*pi*180/(2*(20+2)*pi*asin(2*x))-0.5)};
\addlegendentry[]{Theoretical $q = 20$}
\addplot[
    only marks,
    color=black,
    mark=*,
    error bars/.cd,
    y dir=both,
    y explicit,
    ]%
table[
    x=x,
    y=y,
    y error plus=y-max,
    y error minus=y-min,
    ]{data/q2_k.dat};
\addlegendentry[]{Empirical $q = 2$}

\addplot[
    only marks,
    color=black!33!white,
    mark=diamond*,
    error bars/.cd,
    y dir=both,
    y explicit,
    ]%
table[
    x=x,
    y=y,
    y error plus=y-max,
    y error minus=y-min,
    ]{data/q20_k.dat};
\addlegendentry[]{Empirical $q = 20$}
\end{axis}
\end{tikzpicture}
\begin{tikzpicture}
    \begin{axis}[
    xmajorgrids=true,
    ymajorgrids=true,
    width=0.49\textwidth,
    legend pos=north west,
    legend style={nodes={scale=0.85, transform shape}},
    xmode = log,
    x dir=reverse,
    xlabel=$\epsilon$,
    ylabel=$I$,
    ]
\addplot [
    domain=0.00001:0.1, 
    samples=100, 
    color=black,
    line width = 2pt,
]
{ln(2*2*pi/(2*(2+2)*rad(asin(2*x))))/ln(2)};
\addlegendentry[]{Theoretical $q = 2$}
\addplot [
    domain=0.00001:0.1, 
    samples=100, 
    color=black!33!white,
    line width = 2pt,
    style=densely dotted,
]
{ln(20*20*pi/(2*(20+2)*rad(asin(2*x))))/ln(20)};
\addlegendentry[]{Theoretical $q = 20$}
\addplot[
    only marks,
    color=black,
    mark=*,
    error bars/.cd,
    y dir=both,
    y explicit,
    ]%
table[
    x=x,
    y=y,
    y error plus=y-max,
    y error minus=y-min,
    ]{data/q2_I.dat};
\addlegendentry[]{Empirical $q = 2$}

\addplot[
    only marks,
    color=black!33!white,,
    mark=diamond*,
    error bars/.cd,
    y dir=both,
    y explicit,
    ]%
table[
    x=x,
    y=y,
    y error plus=y-max,
    y error minus=y-min,
    ]{data/q20_I.dat};
\addlegendentry[]{Empirical $q = 20$}
\end{axis}
\end{tikzpicture}
\caption{In the left figure we plot the amplification of the last iteration $k_I$ versus the target precision $\epsilon$. In the right figure we plot the number of iterations $I$ versus the target precision $\epsilon$. The left picture is in a log-log scale with the $x$-axis inverted. The right picture is in a $x-$log scale with the $x$-axis inverted. Each experimental point (circles and diamonds) represent the mean for the last amplification and number of rounds of $100$ experiments. The error bars stretch from the maximum to the minimum value obtained in the same $100$ experiments. The lines are the corresponding theoretical bounds from Equation \eqref{eq:depth_bound} and Equation \eqref{eq:max_iterations_bound}. To avoid clutter, we have omitted the results for $q =10$.}
\label{fig:experimental_k}
\end{figure}
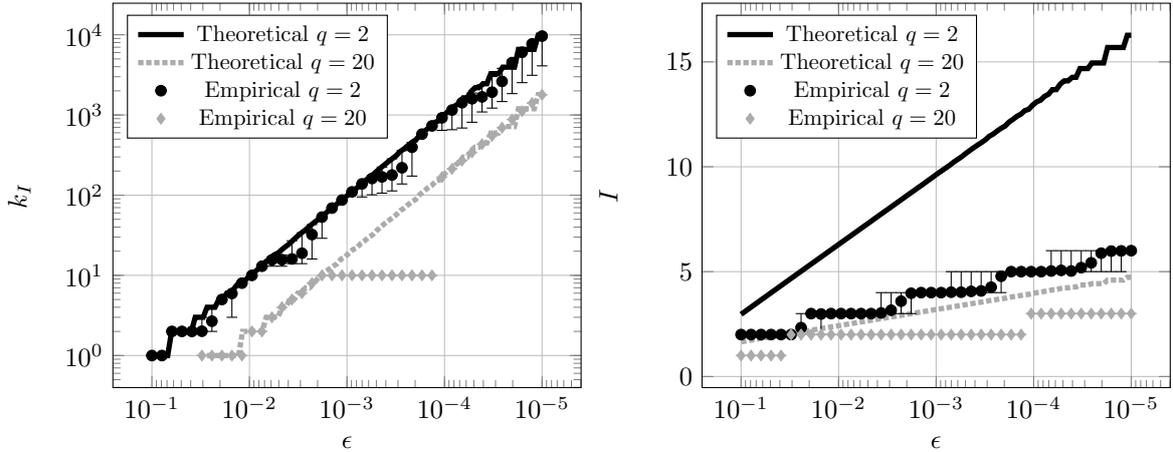
Figure \ref{fig:experimental_k} shows the empirical result and the theoretical bounds for the amplification in the last iteration and the number of iterations performed. In order to explain the behavior depicted in Figure \ref{fig:experimental_k}, we focus on the results obtained for $q = 20$ (grey diamonds). The first five points in the right plot correspond to large values of $\epsilon$ (thus a low target precision); they are associated to the lowest value for the number of iteration $I$, namely $I = 1$. Equivalently, they correspond to $2 k_I + 1 = I = 1$, thus $k_I = 0$. The same five points are not represented in the left plot, because the log scale does not allow to represent $k_I=0$. This phenomenon occurs when the precision of the first estimation, $\epsilon^a_1$, is already lower than the target precision $\epsilon$, so that the algorithm does not need to perform more than one iteration.\\

\noindent 
Following the $q=20$ grey diamonds towards smaller $\epsilon$ in the right plot of Figure \ref{fig:experimental_k}, we note a jump from $I=1$ to $I=2$. For the same points that in the right plot give rise to the $I=2$ plateau, in the left plot we observe two different regimes, one approximately raising linearly, the other being a plateau at $k_I = 10$. In the linearly raising regime, we see that the amplification matches that predicted by the theoretical bound for the maximum amplification $k_I^{\max}$, the experimental diamonds and the theoretical line superpose. Such linear ramp, however, saturates to $k_I = 10$. To give an explanation for this behavior we can proceed as follows. We first compute which is the minimum  possible value for $k_2$ when $q= 20$, namely
\begin{equation}\label{qset}
    \dfrac{2k_2+1}{2k_1+1} = 2k_2+1 = q_1\geq 20\implies k_2 \geq 9.5\ ,
\end{equation}
obtained for $k_1=0$.
Since $k_2$ can only take natural values, we conclude that $k_2$ is at least $10$. Now the explanation of the two regimes becomes more clear. With the first iteration the algorithm does not reach the target precision $\epsilon$, so it proceeds to a second iteration. In this second iteration, as $q=20$, we are sure that $k_2$ is at least $10$. But $k_2=10$ corresponds to too big an amplification, thus the condition in Algorithm \ref{alg:RQAE} on the maximum amplification is activated, thus making the value of the last amplification match the theoretical maximum $k_I = k_I^{\max}$. This occurs until we reach $k_2 = 10$. At this point, $k_I^{\max}$ exceeds the number of amplifications set by $q$ through \eqref{qset}, and the condition on the maximum amplification deactivates. The number of amplifications corresponding to the plateau at $k_I=10$ is by large sufficient to attain the target precision, so that the algorithm maintains the same level of amplification and the same number of iterations for a while, until $k_I=10$ is not enough to obtain the target precision any more. At such point, the algorithm needs to perform a further iteration (the third) and the number of amplifications of the last iterations is again fixed to $k_I^{\max}$.

\section{Conclusion}\label{sec:conclusion}

Throughout this paper we have proposed a new methodology for estimating an amplitude encoded in a quantum circuit. This methodology depends on the possibility of defining a new class of shifted oracles which is easily built, at least in some cases of practical interest. 
For a suitable choice of parameters described in Algorithm \ref{alg:RQAE}, we have proven a set of interesting properties. They include a proof of correctness, a bound on the maximum depth of the algorithm and an upper bound on the total number of shots, hence characterizing the performance of the algorithm.\\

If we compare RQAE to similar algorithms in the literature such as QAE, QAES and IQAE we have three advantages and one caveat. The caveat is obviously that we are constrained by the possibility of constructing an appropriate oracle including the shift. The first advantage is that we are extracting more information from the quantum circuit than just the module of the amplitude. This feature can be of extreme importance for certain applications where the result can be positive or negative. The second advantage is that we can control to some extent the depth of the circuit, a crucial feature in the current NISQ era. The third advantage is that the total number of calls to the oracle is lower to that QAES and IQAE with an appropriate choice of the amplification policy. It is true that QAES has a better order of convergence, however the constants involved are very large, making the method unfeasible for most values of $\epsilon$ used in practice (see the comparison done in \cite{Grinko_2021}).\\ 

The choice of parameters considered in the present paper is not unique and alternative choices could be more efficient in terms, for example, of the total number of shots or the total circuit depth.%
\footnote{A recent study on amplitude estimation with shallow circuits has been performed in \cite{Plekhanov2022variationalquantum}.}
In fact, we believe it is interesting to explore different parameter settings in the future. One such possibility consists in considering to change the number of shots on each iteration in a dynamical manner. This is motivated by the fact that, in the early stages, $q$ has to be large in order to ensure that we get amplification while, as we increase the number of amplifications, we can lower the value of $q$ while still getting amplification. In this sense, we could consider our current proposal as a ``static scheme'' which could be generalized to a ``dynamical'' or ``adaptive'' scheme. Moreover, since we observed that lower values of $q$ tend to be better in terms of the total number of calls to the oracle, a suitable dynamical strategy could further improve the performance of the method.
One could even pose a more ambitious question pursuing the scheme that minimizes the total number of oracle calls. The generalization to a dynamical scheme is not particularly difficult neither from the conceptual nor from the implementation viewpoint. Nonetheless, the proof of theoretical rigorous bounds becomes more challenging than the static scheme considered here.\\ 

Another interesting direction to extend the RQAE algorithm presented here would be that of retrieving not only the sign but the full phase of a complex amplitude. Morally, we would then move in the opposite direction with respect to the standard QAE. Namely, instead of using the QPE to perform a QAE, we would use a QAE algorithm to define an alternative QPE algorithm. Such an extension would probably be challenging in terms of proving rigorous bounds for the performance of the algorithm.

\section{Acknowledgements}
All  authors  acknowledge  the  European  Project  NExt  ApplicationS  of  Quantum  Computing (NEASQC), funded by Horizon 2020 Program inside the call H2020-FETFLAG-2020-01(Grant Agreement 951821).
A. Manzano and Á. Leitao wish to acknowledge the support received from the Centro de Investigación de Galicia ``CITIC", funded by Xunta de Galicia and the European Union (European Regional Development Fund- Galicia 2014-2020 Program), by grant ED431G 2019/01.
D. Musso acknowledges support from the Physics Department at Oviedo University (UNIOVI), the High Energy Physics Group (FPAUO) and the Institute of Space Sciences and Technologies of Asturias (ICTEA).
Part of the computational resources for this project were provided by the Galician Supercomputing Center (CESGA).\\

We would like to thank Elías Combarro, Vedran Dunjko, Andrés Gómez, Javier Mas, María R. Nogueiras, Gustavo Ordóñez, Juan Santos Suárez and Carlos Vázquez for fruitful discussions on some aspects of the present work.

\appendix
\section{Proof of theorem \ref{the:properties}}\label{sec:proof}
Here we proof each of the statements of Theorem \ref{the:properties} one by one in order of appearance.
\subsection{First proposition}
When finding an empirical estimate $\hat{p}$ of a probability $p$, we can assign to it a confidence interval (\emph{i.e.} estimating an associated statistical error) by using Hoeffding's inequality \cite{hoeffding}:%
\footnote{Although there exist tighter bounds, they are much less tractable from an analytic point of view. One such example is  Clopper-Pearson \cite{clopper_pearson}.}
\begin{equation}\label{eqn:Hoeffding_bounds}
    \mathbb{P}\Big(\left|p-\hat{p}\right|\geq {\epsilon}_p \Big)\leq 2e^{-n \epsilon_p^2} = \gamma\ ,
\end{equation}
where $\epsilon_p$ is the precision, $1-\gamma$ is the confidence level and $n$ is the number of shots (\emph{i.e.} samplings) used for the measurement. As we fixed the values for $N_i$ and $\gamma_i$ in Equations \eqref{eq:number_shots} and \eqref{eq:max_confidence}, using \eqref{eqn:Hoeffding_bounds} we get a fixed value for $\epsilon^p_i$:
\begin{equation}
     \mathbb{P}\Big[\left|\sin^2\left[(2k_i+1)\theta_i\right]-\hat p_i\right|\geq\epsilon^p_i\Big] \leq 2e^{-2N_i(\epsilon^p_i)^2} = \gamma_i\ .
\end{equation}
Rewriting the previous expression in terms of $\epsilon^p_i$ we have:
\begin{equation}
    \left(\epsilon^p_i\right)^2 = \dfrac{1}{2N_i}\log\left(\dfrac{2}{\gamma_i}\right) = \dfrac{1}{2\left\lceil N \right\rceil}\log\left(\dfrac{2}{\gamma_i}\right)\leq \dfrac{1}{2N}\log\left(\dfrac{2}{\gamma_i}\right)= (\epsilon^p)^2\ ,
\end{equation}
where we recalled the definition in \eqref{eq:number_shots}. We have thus proven the first proposition.

\subsection{Second proposition}

By definition we have that:
\begin{equation}
    q_i = \dfrac{2k_{i+1}+1}{2k_i+1}\ .
\end{equation}
From Equation \eqref{eq:next_k}, this expression can be rewritten as:
\begin{equation}
    q_i = \dfrac{2\left\lfloor\dfrac{\pi}{4\arcsin(2\epsilon^a_i)}-\dfrac{1}{2}\right\rfloor+1}{2k_i+1}\ .
\end{equation}
We now consider the fact that $\left\lfloor x\right\rfloor\geq x-1$ thus obtaining:
\begin{equation}
    q_i \geq \dfrac{\dfrac{\pi}{2\arcsin(2\epsilon^a_i)}-2}{2k_i+1} = \dfrac{\pi}{2\arcsin(2\epsilon^a_i)(2k_i+1)}-\dfrac{2}{2k_i+1}\ .
\end{equation}
Since $k_i\geq 0$, we have also
\begin{equation}\label{eq:intermediate_step}
    q_i \geq \dfrac{\pi}{2\arcsin(2\epsilon^a_i)(2k_i+1)}-2\ .
\end{equation}
Now we focus on the term $(2k_i+1)\arcsin(2\epsilon_i^a)$ which can be rewritten in terms of $\epsilon^p_i$ as
\begin{equation}\label{eq:ampi_k}
    (2k_{i}+1)\arcsin\left[ \sin\left(\dfrac{\arcsin\left(\sqrt{\min(\hat{p}_i+\epsilon^p_i,1)}\right)}{2k_{i}+1}\right)-\sin\left(\dfrac{\arcsin\left(\sqrt{\max(\hat{p}_i-\epsilon^p_i,0)}\right)}{2k_{i}+1}\right)\right]\ ,
\end{equation}
where we have used \eqref{eqn:following_amplitude_bounds} and \eqref{pmaxmin} after an obvious relabelling of the index.
Next, we define the following functions:
\begin{align}\label{funciona}
    f(\epsilon^p_i;k_i,p_i) :=  (2k_{i}+1)\arcsin&\left[ \sin\left(\dfrac{\arcsin\left(\sqrt{\min(\hat{p}_i+\epsilon^p_i,1)}\right)}{2k_{i}+1}\right)\right.\\ \nonumber &\qquad \qquad \left.-\sin\left(\dfrac{\arcsin\left(\sqrt{\max(\hat{p}_i-\epsilon^p_i,0)}\right)}{2k_{i}+1}\right)\right]\ ,
\end{align}
and
\begin{equation}
    \overline{f}(\epsilon^p_i) :=\arcsin\left(\sqrt{\min(2\epsilon^p_i,1)}\right)\ .
\end{equation}
The function $\overline{f}$ is useful because:
\begin{equation}\label{dise1}
    \overline{f}(\epsilon^p_i) \geq f(\epsilon^p_i;k_i,p_i)\ , \qquad \forall k\in \mathbb{N},\ \hat{p}_i\in [0,1]\ ,
\end{equation}
and we are going to use it to bound expression \eqref{eq:intermediate_step}:
\begin{equation}
     q_i \geq \dfrac{\pi}{2\arcsin(2\epsilon^a_i)(2k_i+1)}-2\geq \dfrac{\pi}{2\arcsin\left(\sqrt{\min(2\epsilon^p,1)}\right)}-2\ .
\end{equation}
As $\epsilon^p_i\leq \epsilon^p\leq \dfrac{1}{2}$ then 
\begin{equation}
     q_i \geq \dfrac{\pi}{2\arcsin\left(\sqrt{\min(2\epsilon^p,1)}\right)}-2\geq \dfrac{\pi}{2\arcsin\left(\sqrt{2\epsilon^p}\right)}-2\ .
\end{equation}
By the definition of $\epsilon^p$ we have that:
\begin{equation}
    q_i\geq q\ ,
\end{equation}
and we have proven the second proposition.
So far, we have not treated the first iteration, $i= 1$, which we now consider explicitly:
\begin{equation}\label{eq:intermediate_first_iteration}
    q_1 = \dfrac{2k_2+1}{2k_1+1} = 2k_2+1 = 2\left\lfloor\dfrac{\pi}{4\arcsin(2\epsilon^a_1)}-\dfrac{1}{2}\right\rfloor+1 \geq \dfrac{\pi}{2\arcsin(2\epsilon^a_1)}-2\ ,
\end{equation}
where we have recalled that $k_1 = 0$.
We focus our attention on the term
\begin{equation}
    \arcsin(2\epsilon^a_1) = \arcsin\left[ \min\left(\dfrac{\hat{p}_{\text{sum}}-\hat{p}_{\text{diff}}}{4b_1}+\dfrac{\epsilon^{p}_1}{|2b_1|},1\right)
    -\max\left(\dfrac{\hat{p}_{\text{sum}}-\hat{p}_{\text{diff}}}{4b_1}-\dfrac{\epsilon^{p}_1}{|2b_1|},-1\right)\right]\ ,
\end{equation}
where we have considered \eqref{eq:first_amplitude_bounds}. 
Following the same strategy as before we define:
\begin{equation}\label{sec:b_function}
    f(\epsilon^p_1) = \arcsin\left[ \min\left(\dfrac{\hat{p}_{\text{sum}}-\hat{p}_{\text{diff}}}{4b_1}+\dfrac{\epsilon^{p}_1}{|2b_1|},1\right)
    -\max\left(\dfrac{\hat{p}_{\text{sum}}-\hat{p}_{\text{diff}}}{4b_1}-\dfrac{\epsilon^{p}_1}{|2b_1|},-1\right)\right]\ .
\end{equation}
An upper bound for \eqref{sec:b_function} is:
\begin{equation}\label{sec:b_function_upper}
    \overline{f}(\epsilon^p_1) = \arcsin\left( \dfrac{\epsilon^p_1}{\left|b_1\right|}\right)\geq f(\epsilon_1^p)\ ,
\end{equation}
which can be directly obtained from \eqref{sec:b_function}.
Hence, from Equation \eqref{eq:intermediate_first_iteration}, we have that:
\begin{equation}\label{dise2}
    q_1 \geq \dfrac{\pi}{2\arcsin\left(\dfrac{\epsilon^p}{|b_1|}\right)}-2\ ,
\end{equation}
where we have used 
\begin{equation}
    2\epsilon^a_1 \geq \frac{\epsilon_1^p}{|b_1|}\ ,
\end{equation}
derived from \eqref{eq:first_amplitude_bounds}.

Eventually, by the definition of $b_1$ we have:
\begin{equation}
    q_1\geq q\ .
\end{equation}
\subsection{Third proposition}

Using \eqref{eqn:following_amplitude_bounds}, we have
\begin{equation}
    \epsilon^a_I(k_I^{\max}) = \frac{1}{2}\sin\left(\dfrac{\arcsin\left(\sqrt{p^{\max}_{I}}\right)}{2k_{I}^{\max}+1}\right)-\frac{1}{2}\sin\left(\dfrac{\arcsin\left(\sqrt{p^{\min}_{I}}\right)}{2k_{I}^{\max}+1}\right)\ .
\end{equation}
Following the same reasoning as in Equation \eqref{target_bou} we have that:
\begin{equation}
    \epsilon^a_I(k_I^{\max}) \leq \dfrac{1}{2}\text{sin}\left(\dfrac{\text{arcsin}\left(\sqrt{2\epsilon^p_I}\right)}{2k_I^{\max}+1}\right)\leq  \dfrac{1}{2}\text{sin}\left(\dfrac{\text{arcsin}\left(\sqrt{2\epsilon^p}\right)}{2\left\lceil\dfrac{1}{2}\dfrac{\arcsin\left(\sqrt{2\epsilon^p}\right)}{\arcsin(2\epsilon)}-\dfrac{1}{2}\right\rceil+1}\right) \leq \epsilon\ ,
\end{equation}
where we have used the first proposition and the definition of $k_I^{\max}$. We have proven the fifth proposition.\\
\subsection{Fourth propositon}
In this subsection we bound the maximum number of iterations needed to achieve the target accuracy $\epsilon$. First note that, if $I$ represents the last iteration, we have that
\begin{equation}\label{penul}
    \epsilon<\epsilon^a_{I-1} =  \frac{1}{2}\sin\left(\dfrac{\arcsin\left(\sqrt{p^{\max}_{I-1}}\right)}{2k_{I-1}+1}\right)-\frac{1}{2}\sin\left(\dfrac{\arcsin\left(\sqrt{p^{\min}_{I-1}}\right)}{2k_{I-1}+1}\right),
\end{equation}
otherwise we would be in the last iteration, and that is false by hypothesis. To write \eqref{penul} we have used \eqref{eqn:following_amplitude_bounds} with $I=i+1$. Using similar arguments as in the previous section, we bound $\epsilon^a_{I-1}$ by
\begin{equation}\label{target_bou}
    \epsilon < \epsilon^a_{I-1} \leq \frac{1}{2}\sin\left(\dfrac{\arcsin\left(\sqrt{2\epsilon^p_{I-1}}\right)}{2k_{I-1}+1}\right)\ .
\end{equation}
We can rewrite \eqref{target_bou} as
\begin{equation}\label{rt2}
\begin{aligned}
   &(2k_1+1)\prod_{i = 1}^{I-2} q_i = 2k_{I-1}+1 <\dfrac{\arcsin\left(\sqrt{2\epsilon^p_{I-1}}\right)}{\arcsin(2\epsilon)}\\
   &\leq \dfrac{\arcsin\left(\sqrt{2\epsilon^p}\right)}{\arcsin(2\epsilon)} =: (2k_1+1)\,\prod_{i = 1}^{T-2}q = (2k_1+1) q^{T-2}\ ,
\end{aligned}
\end{equation}
where we have used $\epsilon^p_{i}\leq \epsilon^p$ and we have introduced the positive number $T$. Still from \eqref{rt2}, we obtain that
\begin{equation}\label{eq:max_iterations_aux}
    \prod_{i = 1}^{I-2} q_i<q^{T-2}\ .
\end{equation}
Using the second proposition $q_i\geq q$ we get
\begin{equation}
     T = 
    \log_q\left(q^2\dfrac{\arcsin\left(\sqrt{2\epsilon^p}\right)}{\arcsin(2\epsilon)}\right)>I.
\end{equation}
This means that, eventually, we have an upper bound $T$ for the number of iterations $I$, when --for each new iteration-- we increase the amplification $K_i=2k_i+1$ by at least a factor $q$
and we have proven the third proposition. Moreover, from \eqref{eq:max_iterations_aux} and $q_i\geq q$ we have that:
\begin{equation}\label{eq:max_iterations_inequality}
    \prod_{i = 1}^{I-2-j} q_i<q^{T-2-j}\ .
\end{equation}
\subsection{Fourth proposition}
We now want to ensure that the precision $\epsilon$ is met with confidence $1-\gamma$. In order to achieve this, note that:
\begin{align*}
     \mathbb{P}\Big[a\not\in [a^{\min}_I,a^{\max}_I]\Big] &=\mathbb{P}\Big[\sin^2\left[(2k_I+1)\theta_{I}\right]\not\in [p^{\min}_I,p^{\max}_I]\Big]\\
     &\leq \mathbb{P}\Big[\bigcup\limits_{i = 1}^{I} \sin^2\left[(2k_i+1)\theta_{i}\right]\not\in [p^{\min}_i,p^{\max}_i]\Big]\\
    &\leq \sum\limits_{i = 1}^I \mathbb{P}\Big[\sin^2\left[(2k_i+1)\theta_{i}\right]\not\in [p^{\min}_i,p^{\max}_i]\Big] \\ 
    &\leq \sum\limits_{i = 1}^I\gamma_i=
    \sum\limits_{i = 1}^I\dfrac{\gamma}{T} = \gamma\dfrac{I}{T}<\gamma ,
\end{align*}
where we have used the definitions and the third proposition. We remind ourselves that $1-\gamma_i$ represents the confidence level of the single iteration.

\subsection{Sixth proposition}
We want to find the necessary maximum number of calls to the oracle in order to obtain a target precision $\epsilon$ with confidence {\color{violet}$1-\gamma$}. Suppose that we finish after $I$ iterations, then the number of calls to the oracle is given by
\begin{equation*}
    N_{\text{oracle}} = \sum_{i = 1 }^{I}N_ik_i =  \sum_{i = 1}^{I}N_ik_i = \sum_{i = 1}^{I}N_i\dfrac{K_i-1}{2},
\end{equation*}
As the number of shots $N_i$ of the individual iteration is constant we have that:
\begin{equation}
    \begin{aligned}
   N_{\text{oracle}} 
   &= \dfrac{N_i}{2}\sum_{i = 1}^{I-1}(K_i-1)+\dfrac{N_iK_I}{2} = \dfrac{N_i}{2}\left(1+\sum_{i = 2}^{I-1}K_i-I\right)+\dfrac{N_iK_I}{2} \\ 
   &=\dfrac{N_i}{2}\left(1+K_1\sum_{i = 1}^{I-2}\prod_{j = 1}^{i}q_i-I\right)+\dfrac{N_iK_I}{2} \\
   &= \dfrac{N_i}{2}\left(1+\sum_{i = 1}^{I-2}\prod_{j = 1}^{i}q_i-I\right)+\dfrac{N_iK_I}{2}\ ,
\end{aligned}
\end{equation}
where we have used $K_1 = 1$, \emph{i.e.} $k_1=0$.
 Then, using the inequality for $q^T$ in \eqref{eq:max_iterations_inequality}, we obtain
\begin{equation}
    N_{\text{oracle}} < \dfrac{N_i}{2}\left(1+\sum_{i = 1}^{I-2}q^{T-I+i}-I\right)+\dfrac{N_iK_I}{2} = \dfrac{N_i}{2}\left(1+q^{T-I+1}\dfrac{1-q^{I-2}}{1-q}-I\right)+\dfrac{N_iK_I}{2},
\end{equation}
where, in order to perform the second step, we have used that $q>1$.
Developing the expression, we have
\begin{align*}
    N_{\text{oracle}} &< \dfrac{N_i}{2}\left(1+\dfrac{q^{T-I+1}}{1-q}-\dfrac{q^{T-1}}{1-q}-I\right)+\dfrac{N_iK_I}{2}\\
    &< \dfrac{N_i}{2}\dfrac{q^{T-1}}{q-1}+\dfrac{N_iK_I}{2}\\&= \dfrac{N_i}{2}q^{T-2}\dfrac{q}{q-1}+\dfrac{N_iK_I}{2}\ ,
\end{align*}
where we have used $I\geq1$ and $\dfrac{q}{q-1}q^{T-I}> 1$.
Eventually, using \eqref{eq:max_iterations_inequality} and the fifth proposition we obtain
\begin{equation}
    N_{\text{oracle}} <
    \dfrac{N_i}{2}K^{\max}\left(1+\dfrac{q}{q-1}\right)\ .
\end{equation}
It is straightforward to define an upper bound for the number of shots:
\begin{flalign*}
    & N_i = \left\lceil N \right\rceil < N+1 = \dfrac{1}{2(\epsilon^p)^2}\log\left(\dfrac{2T}{\gamma}\right)+\log(e) =\\ 
    & = \log\left(\left(\dfrac{2T}{\gamma}\right)^{\dfrac{1}{2(\epsilon^p)^2}}\right)+\log(e) = \log\left(e\left(\dfrac{2T}{\gamma}\right)^{\dfrac{1}{2(\epsilon^p)^2}}\right) = \\
    & = \log\left(\left(e^{2(\epsilon^p)^2}\dfrac{2T}{\gamma}\right)^{\dfrac{1}{2(\epsilon^p)^2}}\right) ={\dfrac{1}{2(\epsilon^p)^2}}\log\left(e^{2(\epsilon^p)^2}\dfrac{2T}{\gamma}\right) < \\
    & < \dfrac{1}{2(\epsilon^p)^2}\log\left(\sqrt{e}\dfrac{2T}{\gamma}\right) =N^{\text{max}}\ ,
\end{flalign*}
Thus, we can have
\begin{equation}
\begin{aligned}
    N_{\text{oracle}}&<  \dfrac{N_i}{2}K_I^{\max}\left(1+\dfrac{q}{q-1}\right)\\
    &<\dfrac{N^{\max}}{2}K_I^{\max}\left(1+\dfrac{q}{q-1}\right)
    =\dfrac{1}{(2\epsilon^p)^2}\log\left(\dfrac{2\sqrt{e}T}{\gamma}\right)K_I^{\max}\left(1+\dfrac{q}{q-1}\right)\ .
    \end{aligned}
\end{equation}

Finally, expressing $\epsilon^p$ in terms of $q$ we get a bound for the number of calls to the oracle in terms of the input parameters $\epsilon$, $\gamma$ and $q$:
\begin{equation}\label{Nq}
    N_{\text{oracle}}<\dfrac{1}{\sin^4\left(\dfrac{\pi}{2(q+2)}\right)}\log\left[\dfrac{2\sqrt{e}\log_q\left(\dfrac{q^2\pi}{2(q+2)\arcsin(2\epsilon)}\right)}{\gamma}\right]\left(\dfrac{\pi}{2(q+2)\arcsin(2\epsilon)}+2\right)\left(1+\dfrac{q}{q-1}\right)\ ,
\end{equation}
and we have proven the sixth proposition.\\ \\
We have done the proof for the case where we consider the number of calls to the oracle as $\sum_{i=1}^I N_ik_i$. This is the same as in IQAE (see \cite{Grinko_2021}). However, here we are only computing the number of calls to the Grover oracle and that cannot be considered a fair comparison with the unamplified case where we are referring to the number of calls to the original oracle $\mathcal{A}$. For that reason we will show what will be the bound for the number of calls to the oracle $\mathcal{A}$ when we consider $\mathbb{A}^\dagger$ the same as $\mathcal{A}$. The number of calls would then be defined as:
\begin{equation}
\begin{aligned}
    N^{\mathcal{A}}_{\text{oracle}} &= N_1+\sum_{i = 2}^{I-1}N_i(2k_i+1)+(2k_I+1)N_I = N_i\left(1+\sum_{i = 2}^{I-1}(2k_i+1)+(2k_I+1)\right) \\
    &= N_i\left(1+\sum_{i = 2}^{I-2}\prod_{j=1}^{i}q_i+(2k_I+1)\right)\ ,
\end{aligned}
\end{equation}
where we have used $2k_i$ because each time you call the Grover oracle you are calling once to $\mathcal{A}$ and once to $\mathcal{A}^{\dagger}$. The term $+1$ is for the first aplication of the oracle (recall that we are calling the operator $\mathcal{A}\mathcal{G}^k$). In the first iteration we don't call to the Grover oracle, so we only $N_i$ calls to $\mathcal{A}$. Following the same reasoning as before we can bound the term $1+\sum_{i = 2}^{I-2}\prod_{j=1}^{i}q_i$
\begin{equation}
\begin{aligned}
    N^{\mathcal{A}}_{\text{oracle}} &= N_i\left(1+\sum_{i = 2}^{I-2}\prod_{j=1}^{i}q_i+(2k_I+1)\right)\\
    &< N_i\left(1+q^{T-2}\dfrac{q}{q-1}+(2k_I+1)\right)\leq  N_i\left(1+\left(1+\dfrac{q}{q-1}\right)K^{\max}\right)\ .
\end{aligned}
\end{equation}
Finally we see that:
\begin{equation}
    N^{\mathcal{A}}_{\text{oracle}}\approx 2N_{\text{oracle}}+N_i\ .
\end{equation}
\section{Constructing the shifted states}\label{sec:shift}

Given an oracle $\mathcal{G}$ such that:
\begin{equation}
    \mathcal{G}\ket{0} = a\ket{0}+\sqrt{1-a^2}\ket{0^\perp},
\end{equation}
we can build an oracle $\mathcal{A}_{\theta_b}$ such that:
\begin{equation}
        \mathcal{A}_{\theta_b}\left|0\right)\ket{0}= \frac{a+b}{2}\left|0\right)\ket{0}+...\ .
\end{equation}
\\ \\
We start by applying a Hadamard gate to the auxiliary register:
\begin{equation}
   \left(H\otimes \mathbb{1}\right)  \left|0\right)\ket{0} =\dfrac{1}{\sqrt{2}}\left( \left|0\right)\ket{0}+\left|1\right)\ket{0}\right).
\end{equation}
Next, apply $\mathcal{G}$ controlled in the auxiliary register:
\begin{equation}
    c\otimes \mathcal{G}\left(\dfrac{1}{\sqrt{2}}\left( \left|0\right)\ket{0}+\left|1\right)\ket{0}\right)\right) = \dfrac{1}{\sqrt{2}}\left( \left|0\right)\ket{0}+a\left|1\right)\ket{0}+\sqrt{1-a^2}\left|1\right)\ket{0^\perp}\right).
\end{equation}
We continue by applying $y$-rotation with angle $\theta_b$ controlled in the auxiliary register: 
\begin{equation}
    \overline{c}\otimes \mathcal{R}_y(\theta_b)\left(\dfrac{1}{\sqrt{2}}\left( \left|0\right)\ket{0}+a\left|1\right)\ket{0}+\sqrt{1-a^2}\left|1\right)\ket{0^\perp}\right)\right) = \dfrac{1}{\sqrt{2}}\left( \cos(\theta_b)\left|0\right)\ket{0}+a\left|1\right)\ket{0}\right)+...\ .
\end{equation}
Finally, applying a Hadamard gate to the first register we get:
\begin{equation}
     \left(H\otimes \mathbb{1}\right)\left(\dfrac{1}{\sqrt{2}}\left( \cos(\theta_b)\left|0\right)\ket{0}+a\left|1\right)\ket{0}\right)+...\right) = \frac{a+\cos(\theta)}{2}\left|0\right)\ket{0}+\frac{-a+\cos(\theta)}{2}\left|1\right)\ket{0}+...\ .
\end{equation}

\section{Estimation of the
probability}\label{sec:probability_vs_amplitude}

In Appendix \ref{sec:proof} we have proven some bounds for the amplitudes. However, in the literature one typically finds analogous bounds proven for the probabilities (\emph{i.e.} for the square of the amplitudes). In this Appendix we show that, under some restrictions, given an estimation for the amplitudes $(a_{\min},a_{\max})$ such that $a_{\max}- a_{\max}\leq 2\epsilon$, we can build a pair of bounds for the probabilities $(p_{\min},p_{\max})$ such that $p_{\max}- p_{\max}\leq 2\epsilon$. So, we can indistinctly refer to the properties of the method when estimating amplitudes or estimating probabilities.
\begin{enumerate}
    \item \textbf{Case 1}: $a_{\max}>0$ and $a_{\min}>0$.\\
    We build the bounds for the probability as $p_{\max} = a_{\max}^2$, $p_{\min} = a_{\min}^2$.
    \begin{equation}
        p_{\max}-p_{\min} = a_{\max}^2-a_{\min}^2 = (a_{\max}+a_{\min})(a_{\max}-a_{\min})\leq (a_{\max}-a_{\min})\leq 2\epsilon,
    \end{equation}
    where we have used that $\left|a_{\max}\right|\leq 0.5\ $, $\left|a_{\min}\right|\leq 0.5\ $.
    \item \textbf{Case 2}: $a_{\max}<0$ and $a_{\min}<0$.\\
    We build the bounds for the probability as $p_{\max} = a_{\min}^2$, $p_{\min} = a_{\max}^2$.
    \begin{equation}
        p_{\max}-p_{\min} = a_{\min}^2-a_{\max}^2 = (-a_{\max}-a_{\min})(a_{\max}-a_{\min})\leq (a_{\max}-a_{\min})\leq 2\epsilon,
    \end{equation}
    where we have used that $\left|a_{\max}\right|\leq 0.5\ \ $, $\left|a_{\min}\right|\leq 0.5\ \ $.
        \item \textbf{Case 3}: $a_{\max}>0$ and $a_{\min}<0$ and $|a_{\max}|>|a_{\min}|$\\
    We build the bounds for the probability as $p_{\max} = a_{\max}^2$, $p_{\min} = 0$.
    \begin{equation}
        p_{\max}-p_{\min} = a_{\max}^2\leq a_{\max}-0\leq a_{\max}-a_{\min} \leq 2\epsilon.
    \end{equation}
            \item \textbf{Case 4}: $a_{\max}>0$ and $a_{\min}<0$ and $|a_{\max}|<|a_{\min}|$\\
    We build the bounds for the probability as $p_{\max} = a_{\min}^2$, $p_{\min} = 0$.
    \begin{equation}
        p_{\max}-p_{\min} = a_{\min}^2\leq |a_{\min}|\leq a_{\max}-a_{\min} \leq 2\epsilon.
    \end{equation}
\end{enumerate}

\printbibliography

\end{document}